\documentclass{tdp}

\usepackage[utf8]{inputenc} % For weird characters in bib :)
\usepackage[english]{babel}

\usepackage[svgnames,table]{xcolor}

\usepackage{amsmath}

\newtheorem{mydef}{Definition}
\usepackage{enumerate}
\usepackage{enumitem} % Prevent overlap in description environment

\usepackage{balance}

\usepackage{amssymb}

\usepackage{booktabs}%midrule etc

\usepackage{tikz}
\usepackage{siunitx}
\usetikzlibrary{arrows,decorations,backgrounds,patterns,matrix,shapes,fit,calc,shadows,plotmarks}
\usetikzlibrary{arrows,decorations.pathmorphing,backgrounds,fit,positioning,shapes.symbols,chains}

\usetikzlibrary{patterns,shadows}
\usetikzlibrary[topaths]

\usepackage{url}
\usepackage{hyperref} %Clickable links and references
\usepackage{xspace}
\usepackage{graphicx} %to use includegraphics command
\usepackage{lipsum}
\usepackage{array}
\usepackage{tabularx}
\usepackage{multirow}
\usepackage{ctable}
\usepackage{booktabs}
\usepackage{threeparttable}
\usepackage[draft]{todonotes} %check for the compile time
\usepackage{xtab}% Multipage tables, longtable/supertabular doesn't work for double column...

\usepackage{subcaption}
\usepackage{float}
\usepackage[section]{placeins} % Floatbarriers for sections!

\usepackage{appendix}
\usepackage{longtable}

\usepackage{enumitem}
\setdescription{leftmargin=0pt}

\usepackage{csquotes}
\makeatletter
\let\c@author\relax
\makeatother
\usepackage[sorting=none, style=numeric-comp, natbib=true,maxcitenames=2, mincitenames=1, maxbibnames=99, date=year, url=false, doi=false, isbn=false, backend=biber, giveninits=true]{biblatex}
\usepackage{forest}

\definecolor{blue1}{RGB}{222,235,247}
\definecolor{blue2}{RGB}{158,202,225}
\definecolor{blue3}{RGB}{49,130,189}

\newcolumntype{K}{>{\raggedleft\arraybackslash}X}
\newcolumntype{Z}{>{\raggedright\arraybackslash}X}

\definecolor{light-gray}{gray}{0.95}
\definecolor{medium-gray}{gray}{0.70}
\definecolor{dark-gray}{gray}{0.50}

\DeclareCaptionFont{blue}{\color{blue3}}
\newcommand{\foo}{\color{blue2}\makebox[0pt]{\textbullet}\hskip-0.5pt\vrule width 1pt\hspace{\labelsep}}
\usepackage{colortbl}

\usepackage{cleveref}
\crefname{mydef}{Definition}{Definitions}
\crefname{appsec}{Appendix}{Appendices}

\usepackage{mfirstuc}

\begin{document}
% Instructions
% Final version of the paper {\bf should be camera ready} using tdp.cls style file. This is an open access journal with no publishing costs for the authors. The {\em editorial office} is extremely tiny (one person!), and we expect to apply only some minor changes (e.g. pagination) to the final file. So, please adhere completely to the style of the journal. 

% Important: Avoid any change in the margins of the paper. Avoid figures and tables that do not fit the page size. 

% Please, all figures should be in a single type of file. If possible, .eps or .pdf. 

% Please, take special care of uniform capitalization in titles. 

% Please, use uniform style for references. We do not edit them. Bibtex is not required but if used, supply appropriate files. 

% Use standard way for citations~\cite{ref:Author.2008}. 

% Figures and tables following standard latex style. Captions below figures and tables. 

% \section*{Acknowledgements}
% Acknowledgements go here. Partial support by A.B.C is acknowledged. 

\title{SoK: Chasing Accuracy and Privacy, and Catching Both in Differentially Private Histogram Publication}
\author{Boel Nelson$^{*\dagger}$, Jenni Reuben$^{*\ddagger}$}
\address{$^{\dagger}$Department of Computer Science and Engineering, Chalmers University of Technology, SE-412 96 Gothenburg, Sweden. \\ 
  $^{\ddagger}$Department of Mathematics and Computer Science, Karlstad University, SE-651 88 Karlstad, Sweden.\\
  E-mail: {\small \tt{boeln@chalmers.se}}, {\small \tt{jenni.reuben@kau.se}}
}
%% Avoid if possible
\thanks{Both the authors contributed substantially, and share first authorship. The names are ordered alphabetically.}%
%% This part here is not needed for submission
\TDPRunningAuthors{Boel Nelson, Jenni Reuben}
\TDPRunningTitle{Chasing Accuracy and Privacy, and Catching Both}
% \TDPThisVolume{1}
% \TDPThisYear{2008}
% \TDPFirstPageNumber{1000}
% \TDPSubmissionDates{Received 29 February 2008; received in revised form 30 February 2008; accepted 1 March 2008}

\maketitle

\newcommand{\x}{$\checkmark$}
\newcommand{\singlequerysymbol}{$\bullet$}
\newcommand{\workloadquerysymbol}{$\cdots$}
\newcommand{\syntheticdatasymbol}{$\blacksquare$}
\newcommand{\offlineprocessingsymbol}{$\star$}
\newcommand{\postprocessingsymbol}{$\dag$}
\newcommand{\boundedsymbol}{$\clubsuit$}
\newcommand{\unboundedsymbol}{$\diamondsuit$}
\newcommand{\correlatedsymbol}{$\Join$}
\newcommand{\staticsymbol}{$\bar{x}$}
\newcommand{\dynamicsymbol}{$\vec{x}$}
\newcommand{\sparsesymbol}{$\odot$}

\newcommand{\dataset}{data set}
\newcommand{\tradeoff}{trade off}
\newcommand{\mathepsilon}{\varepsilon}
\newcommand{\ep}{$\mathepsilon$}
\newcommand{\edDP}{(\ep, $\delta$)-differential privacy}
\newcommand{\eddp}{(\ep, $\delta$)-DP}
\newcommand{\eDP}{\ep-differential privacy}
\newcommand{\edp}{\ep-DP}
\newcommand{\wevent}{$\mathit{w}$-event level privacy}
\newcommand{\event}{event level privacy}
\newcommand{\Event}{Event level privacy}
\newcommand{\user}{user level privacy}
\newcommand{\User}{User level privacy}
\newcommand{\nodeDP}{Node-DP}
\newcommand{\edgeDP}{Edge-DP}
\newcommand{\highdimension}{high-dimension}
\newcommand{\preprocessing}{pre-processing}
\newcommand{\postprocessing}{post-processing}
\newcommand{\Preprocessing}{Pre-processing}
\newcommand{\Postprocessing}{Post-processing}

\newcommand{\efpa}{{\footnotesize{EFPA}}}
\newcommand{\php}{{\footnotesize{P-HP}}}
\newcommand{\dpcocgen}{{\footnotesize{DPCocGen}}}
\newcommand{\pmost}{{\footnotesize{PMost}}}
\newcommand{\bmax}{{\footnotesize{BMax}}}
\newcommand{\thetaomegahistogram}{{\footnotesize{($\theta$,$\Omega$)-Histogram}}}
\newcommand{\thetacumhisto}{{\footnotesize{$\theta$-CumHisto}}}
\newcommand{\sortaki}{{\footnotesize{SORTaki}}}
\newcommand{\tlambda}{{\footnotesize{T\textsuperscript{$\lambda$}}}}
\newcommand{\prish}{{\footnotesize{PriSH}}}
\newcommand{\gga}{{\footnotesize{GGA}}}
\newcommand{\boost}{{\footnotesize{Boost}}}
\newcommand{\outlierhistopub}{{\footnotesize{Outlier-HistoPub}}}
\newcommand{\pythia}{{\footnotesize{Pythia}}}
\newcommand{\delphi}{{\footnotesize{Delphi}}}
\newcommand{\citm}{{\footnotesize{C$i$TM}}}
\newcommand{\dpcopula}{{\footnotesize{DPCopula}}}
\newcommand{\admm}{{\footnotesize{ADMM}}}
\newcommand{\dsat}{{\footnotesize{DSAT}}}
\newcommand{\dsft}{{\footnotesize{DSFT}}}
\newcommand{\ihp}{{\footnotesize{IHP}}}
\newcommand{\mihp}{{\footnotesize{mIHP}}}
\newcommand{\rcf}{{\footnotesize{RCF}}}
\newcommand{\ac}{{\footnotesize{AC}}}
\newcommand{\drpp}{{\footnotesize{DRPP}}}
\newcommand{\pegs}{{\footnotesize{PeGS}}}
\newcommand{\pegsrs}{{\footnotesize{PeGS.rs}}}
\newcommand{\bpm}{{\footnotesize{BPM}}}
\newcommand{\tru}{{\footnotesize{Tru}}}
\newcommand{\minalg}{{\footnotesize{Min}}}
\newcommand{\opt}{{\footnotesize{Opt}}}
\newcommand{\privelet}{{\footnotesize{Privelet}}}
\newcommand{\priveletplus}{{\footnotesize{Privelet$^+$}}}
\newcommand{\priveletstar}{{\footnotesize{Privelet$^*$}}}
\newcommand{\dpcube}{{\footnotesize{DPCube}}}
\newcommand{\noisefirst}{{\footnotesize{NF}}}
\newcommand{\structurefirst}{{\footnotesize{SF}}}
\newcommand{\dppro}{{\footnotesize{DPPro}}}
\newcommand{\ahp}{{\footnotesize{AHP}}}
\newcommand{\privbayes}{{\footnotesize{PrivBayes}}}
\newcommand{\privtree}{{\footnotesize{PrivTree}}}
\newcommand{\rg}{{\footnotesize{RG}}}

\newcommand{\nieutilityoptimized}{{\footnotesize{NYHXZW19}}}
\newcommand{\cheniterative}{{\footnotesize{CWCW13}}}
\newcommand{\hayboosting}{{\footnotesize{HRMS10}}}
\newcommand{\benkhelifcoclustering}{{\footnotesize{BFCR17}}}
\newcommand{\lidifferential}{{\footnotesize{LYSZ17}}}
\newcommand{\xiaodifferential}{{\footnotesize{XWG11}}}
\newcommand{\dingdifferentially}{{\footnotesize{DWHL11}}}
\newcommand{\wangdifferentially}{{\footnotesize{WGXLY17}}}
\newcommand{\xudifferentially}{{\footnotesize{XZXYYW13}}}
\newcommand{\lidifferentially}{{\footnotesize{LXJL15}}}
\newcommand{\yandifferentially}{{\footnotesize{YZLLLS18}}}
\newcommand{\acsdifferentially}{{\footnotesize{ACC12}}}
\newcommand{\lixiongdifferentially}{{\footnotesize{LXJ14}}}
\newcommand{\xiaodpcube}{{\footnotesize{XXFGL14}}}
\newcommand{\xudppro}{{\footnotesize{XRZQR17}}}
\newcommand{\gaodynamic}{{\footnotesize{GM18}}}
\newcommand{\lugenerating}{{\footnotesize{LMG14}}}
\newcommand{\liuhistogram}{{\footnotesize{LL19}}}
\newcommand{\liihp}{{\footnotesize{LCMM19}}}
\newcommand{\leemaximum}{{\footnotesize{LWK15}}}
\newcommand{\parkpegs}{{\footnotesize{PG14}}}
\newcommand{\hadianprivacypreserving}{{\footnotesize{HLAM16}}}
\newcommand{\dingprivacypreserving}{{\footnotesize{DZBJ18}}}
\newcommand{\barakprivacy}{{\footnotesize{BCDKMT07}}}
\newcommand{\chenprivate}{{\footnotesize{CSJ15}}}
\newcommand{\wangprivate}{{\footnotesize{WHWDXY16}}}
\newcommand{\zhangprivbayes}{{\footnotesize{ZCPSX14}}}
\newcommand{\zhangprivtree}{{\footnotesize{ZXX16}}}
\newcommand{\daypublishing}{{\footnotesize{DLL16}}}
\newcommand{\qianpublishing}{{\footnotesize{QLZCYZ18}}}
\newcommand{\hanpublishing}{{\footnotesize{HSLMZD16}}}
\newcommand{\ghanepublishing}{{\footnotesize{GKR18}}}
\newcommand{\kotsogiannispythia}{{\footnotesize{KMHM17}}}
\newcommand{\liresearch}{{\footnotesize{LL18}}}
\newcommand{\doudalissortaki}{{\footnotesize{DM17}}}
\newcommand{\zhangtowards}{{\footnotesize{ZCXMX14}}}

\newcommand{\qardajidifferentially}{{\footnotesize{QYL13-ICDE}}}
\newcommand{\kellarisdifferentially}{{\footnotesize{KPXP14}}}
\newcommand{\lidata}{{\footnotesize{LHMW14}}}
\newcommand{\machanavajjhalaprivacy}{{\footnotesize{MKAGV08}}}
\newcommand{\blockidifferentially}{{\footnotesize{BBDS13}}}
\newcommand{\fanadaptive}{{\footnotesize{FX13}}}
\newcommand{\raskhodnikovaefficient}{{\footnotesize{RS15}}}
\newcommand{\rastogidifferentially}{{\footnotesize{RN10}}}
\newcommand{\kellarispractical}{{\footnotesize{KG13}}}
\newcommand{\qardajiunderstanding}{{\footnotesize{QYL13-VLDB}}}
\newcommand{\jagadishoptimal}{{\footnotesize{JKMPSS98}}}
\newcommand{\chendifferentially}{{\footnotesize{CXZX15}}}
\newcommand{\lioptimizing}{{\footnotesize{LHRMM10}}}
\newcommand{\hardtsimple}{{\footnotesize{HLM12}}}
\newcommand{\chendifferentiallyngram}{{\footnotesize{CAC12}}}
\newcommand{\mcsherryprivacy}{{\footnotesize{M09}}}
\newcommand{\zhangprivgene}{{\footnotesize{ZXYZW13}}}
\newcommand{\rubinsteinlearning}{{\footnotesize{RBHT12}}}
\newcommand{\qardajipriview}{{\footnotesize{QWL14}}}
\newcommand{\cormodedifferentiallypsd}{{\footnotesize{CPSSY12}}}
\newcommand{\erlingssonrappor}{{\footnotesize{EPK14}}}
\newcommand{\kasiviswanathananalyzing}{{\footnotesize{KNRS13}}}
\newcommand{\sudifferentially}{{\footnotesize{SJNEH16}}}
\newcommand{\hardtmultiplicative}{{\footnotesize{HR10}}}
\newcommand{\liadaptive}{{\footnotesize{LM12}}}
\newcommand{\arasudata}{{\footnotesize{AKL11}}}
\newcommand{\chaudhuridifferentially}{{\footnotesize{CMS11}}}
\newcommand{\cormodedifferentially}{{\footnotesize{CPST12}}}
\newcommand{\chendifferentiallytransit}{{\footnotesize{CFDS12}}}
\newcommand{\duncandisclosure}{{\footnotesize{DS04}}}
\newcommand{\xiedistributed}{{\footnotesize{XTK08}}}
\newcommand{\duchilocal}{{\footnotesize{DJW13}}}
\newcommand{\bassilylocal}{{\footnotesize{BS15}}}
\newcommand{\xiaooptimal}{{\footnotesize{XTC09}}}
\newcommand{\inanprivate}{{\footnotesize{IKGB10}}}
\newcommand{\kasiviswanathanwhat}{{\footnotesize{KLNRS11}}}

\begin{abstract}
% Problem statement: Data analysis good, privacy problems bad. How do we continue to use analysis without violating privacy?
Histograms and synthetic data are of key importance in data analysis. However, researchers have shown that even aggregated data such as histograms, containing no obvious sensitive attributes, can result in privacy leakage. To enable data analysis, a strong notion of privacy is required to avoid risking unintended privacy violations.

% Differential privacy to the rescue! Still a problem: accuracy
Such a strong notion of privacy is \textit{differential privacy}, a statistical notion of privacy that makes privacy leakage quantifiable. The caveat regarding differential privacy is that while it has strong guarantees for privacy, privacy comes at a cost of accuracy. Despite this \tradeoff\ being a central and important issue in the adoption of differential privacy, there exists a gap in the literature regarding providing an understanding of the \tradeoff\ and how to address it appropriately. %Consequently, many differentially private algorithms have been proposed to improve accuracy in novel, non-trivial, ways. However, there does not yet exist a conclusive understanding among the research community as far as publications go, as to why these strategies work and how they can be further combined.

Through a systematic literature review (SLR), we investigate the state-of-the-art within accuracy improving differentially private algorithms for histogram and synthetic data publishing. Our contribution is two-fold: 1) we identify trends and connections in the contributions to the field of diffential privacy for histograms and synthetic data and 2) we provide an understanding of the privacy/accuracy \tradeoff\ challenge by crystallizing different dimensions to accuracy improvement. Accordingly, we position and visualize the ideas in relation to each other and external work, and deconstruct each algorithm to examine the building blocks separately with the aim of pinpointing which dimension of accuracy improvement each technique/approach is targeting. Hence, this systematization of knowledge (SoK) provides an understanding of in which dimensions and how accuracy improvement can be pursued without sacrificing privacy.

% \textit{categories} of accuracy improving techniques, identify the \textit{optimization goal} of each category, and 2) we pave the way for future work
%and investigate the orthogonality of the categories to discuss possible composability
%https://ieeecs-media.computer.org/assets/pdf/taxonomy.pdf
%\keywords{Provenance, \pia, privacy risks}
%\keywords{Privacy Impact Assessment, provenance, privacy.}
% \begin{IEEEkeywords}
% Differential Privacy;
% \end{IEEEkeywords} 
\end{abstract}

\iffalse
Why do the research community need a literature survey like this?, for example the following excerpt is taking from a paper titled "SoK: Security and Privacy in Machine Learning" 

There is growing recognition that ML exposes new vulnerabilities in software systems, yet the technical community's understanding of the nature and extent of these vulnerabilities remains limited.

There are already two survey exists in this area of interest, why do the community need another one?

What is my aim, intuitive answer to the question or tentative answer, how did I achieve it (method) and in the process what do I have contributed (what are contributions)
\fi
%
%What it is?
%Why is it relevant?
%What is the solution?
%Where is the evidence?

\begin{keywords}
accuracy improvement, data privacy, differential privacy, dimensionality reduction, error reduction, histogram, noise reduction, sensitivity reduction, synthetic data, SLR, SoK, systematic literature review, systematization of knowledge, taxonomy, utility improvement
\end{keywords}

\section{Introduction}\label{sec:introduction}
% Data analysis = good shit, we want to be able to do it!
%Drawing insights from collections of personal records is a powerful tool in social science and medical research. However, the data analysis tasks must not violate the privacy of the individuals in the \dataset. For example, data shared on social media is a valuable source for a social science researcher who wants to understand how individuals fleeing from war zones organize themselves. However, access to such information jeopardize the privacy of those individuals. The goal of privacy preserving data analysis is to enable analysis yet protecting individuals from privacy-compromising disclosures from the \dataset.

Being able to draw analytical insights from \dataset s about individuals is a powerful skill, both in business, and in research. However, to enable data collection, and consequently data analysis, the individuals' privacy must not be violated. Some strategies~\cite{samarati_protecting_1998, machanavajjhala_l-diversity_2007, li_t-closeness_2007} for privacy-preserving data analysis focus on sanitizing data, but such approaches require identifying sensitive attributes and also does not consider auxiliary information. As pointed out by~\citet{narayanan_myths_2010}, personally identifiable information has no technical meaning, and thus cannot be removed from \dataset s in a safe way. Furthermore, \citet{dwork_differential_2006} proves that for essentially any non-trivial algorithm, there exists auxiliary information that can enable a privacy breach that would not have been possible without the knowledge learned from the data analysis. Consequently, a strong notion of privacy is needed to avoid any potential privacy violations, while still enabling data analysis.

%What is more, differential privacy is resistant to attacks involving auxiliary data~\cite{ganta_composition_2008}.

Such a strong notion of privacy is \textit{differential privacy}~\cite{dwork_calibrating_2006} (\Cref{sec:dp}), which currently is the de-facto standard for private data analysis~\cite{erlingsson_rappor_2014,Abowd_census_DP_2018}. Differential privacy is a privacy model that provides meaningful privacy guarantees to individuals in the \dataset s by quantifying their privacy loss. The power of differential privacy lies in allowing an analyst to learn statistical correlations about a population, while not being able to infer information about any one individual. To this end, a differential private analysis may inject random noise to the results and this approximated results are then released to the analysts.

\begin{figure}[htbp]
    \centering
    \includegraphics[width=\textwidth]{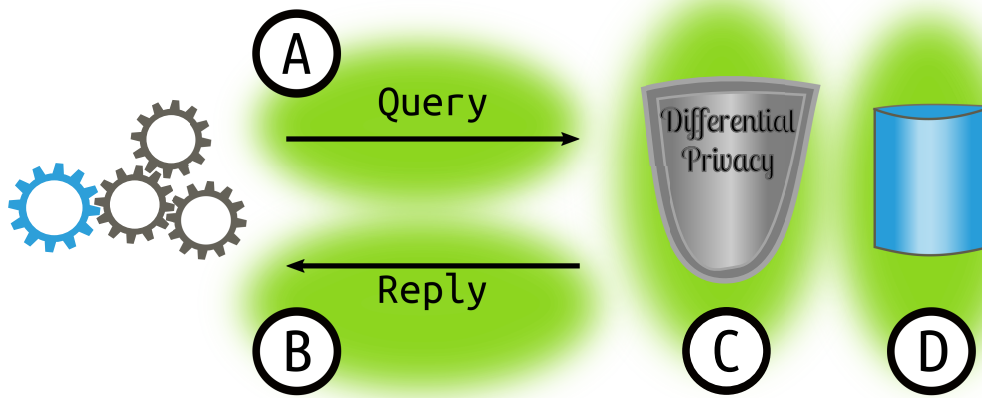}
    \caption{Places for accuracy improvement: A) Altering the query, B) Post-processing, C) Change in the release mechanism, D) Pre-processing}
    \label{fig:improvement-places}
\end{figure}

Differential privacy has spurred a flood of research in devising differentially private algorithms for various data analysis with varying utility guarantees. Given a general workflow of a differentially private analysis, which is illustrated in \Cref{fig:improvement-places}, there exists four \textit{places} (labeled A, B, C and D) for exploring different possibilities to improve accuracy of differential private analyses.

In this work, we focus specifically on differentially private algorithms for histogram, and synthetic data publication. Histograms and synthetic data are particularly interesting because they provide an approximation of the underlying data distribution and synthesis of original data respectively. Intrigued by the idea that there exists several ways to improve accuracy of privatized histograms and synthetic data without compromising privacy, we aim to systematically synthesize the state-of-the-art.

Advancement in research in differentially private histograms and synthetic data publication has received considerable interest within the computer science and statistics research communities~\cite{bowen_comparative_2016,li_lit_survey_2015,meng_different_2017}. However, only a few works systematically and critically assess the state-of-the-art differentially private, accuracy improving algorithms for histograms or synthetic data. \citet{li_lit_survey_2015} and \citet{meng_different_2017} categorized different differentially private publication techniques for both histogram as well as synthetic data and histogram respectively. However, their selection and categorization of the algorithms are not systematic. Further, the selected algorithms in their work are not exclusively accuracy improving techniques, but rather differentially private release mechanisms for histogram and synthetic data. \citet{bowen_comparative_2016}, on the other hand, evaluated using simulation studies, several algorithms for differentially private histogram and synthetic data publication. Their aim is quite different from ours, is to assess the accuracy\footnote{We will use the terms accuracy and utility interchangeably when we refer to decreasing the error, i.e the distance between the privatized result and the true results.} and usefulness\footnote{We use the term usefulness to refer to the impact of the privatized results to conduct statistical inferences.} of the privatized results.

Consequently, to bridge the knowledge gap, the present paper aims to provide a systematization of knowledge concerning differentially private accuracy improving methods for histogram and synthetic data publication. To this end, we qualitatively analyse the literature in this domain using systematic literature review. In order to pave the way for future research, this systematization of knowledge provides a conceptual understanding of enhancing accuracy in the light of privacy constraints.

%we conduct a systematic literature review (SLR) (Section~\Cref{sec:method}) on accuracy improvement techniques for differentially private histogram and synthetic data publication. Our aim with the SLR is: to summarize the state-of-the-art (\Cref{sec:paper-summaries}), identify trends in the literature, and to analyze the constructs of differentially private accuracy improvement strategies, in order to pave the way for future research.

%The validity of selection and categorization of differentially private histogram and synthetic data algorithms included in the literature review by because is not systematically synthesized.
%are not systematic hence scietific reproducability  
% is not systematic identification of 
% and also any histogram or synthetic data publication strategies
%to provide a snapshot of 
% to identify trends  

Our main contributions are:
\begin{enumerate}
    \item A technical summary of each algorithms in order to provide a consolidate view of the state-of-the-art (\Cref{sec:paper-summaries}).
    \item Categorization that synthesize the evolutionary relationships of the research domain in differential privacy for histogram and synthetic data publication (\Cref{sec:positioning}).
    \item Categorization of the state-of-the-art, which is based on the conceptual relationships of the identified algorithms (\Cref{sec:dimensions}).
\end{enumerate}

%Major contributions of our paper 
%Our contribution is a systematically structured, qualitative analysis (\Cref{sec:analysis}) of accuracy improving differentially private histogram or synthetic data publishing algorithms. More concretely, we provide: 
    %\item A discussion on how the different techniques can be composed to build new algorithms where we open up for future work (\Cref{sec:discussion}).

%impact (basically say we fill the gap)
%This systematization of knowledge provides an understanding of accuracy improving techniques as well as an analysis of the efficient combinations of techniques to improve accuracy without sacrificing privacy.

\section{Differential Privacy}\label{sec:dp}
%Intro to DP, quantifying and bounding privacy loss!
%Original definition
Differential privacy~\cite{dwork_calibrating_2006} is a statistical definition that enables privacy loss to be quantified and bounded. In differential privacy, privacy loss is bounded by the parameter \ep\. To achieve trivial accuracy improvement, \ep\ can be tweaked to a higher value, as this gives less privacy (greater privacy loss) which means more accuracy. In this paper we only consider accuracy improvements in settings where \ep\ is fixed.

We formally define \eDP\ in \Cref{def:eDP}, based on \citet{dwork_differential_2006}. The parameter \ep\ is usually referred to as the \textit{privacy budget}. Essentially, \ep\ is the cost in terms of privacy loss for an individual participating in an analysis.

\begin{samepage}
\begin{mydef}[\ep-Differential Privacy]\label{def:eDP}
A randomized algorithm $f'$ gives \eDP\ if for all \dataset s $D_1$ and $D_2$, where $D_1$ and $D_2$ are neighboring, and all $\mathcal{S}$ $\subseteq$ Range($f'$),
\[
\text{Pr}[f'(D_1) \in \mathcal{S}]\leq e^{\mathepsilon} \times \text{Pr}[f'(D_2) \in \mathcal{S}]
\]
\end{mydef}
\end{samepage}

% Also something on (e,d)-DP?
A relaxed version of differential privacy is \edDP~\citet{dwork_our_2006}, which we define in \Cref{def:edDP}. \edDP\ is primarily used to achieve better accuracy, but adds a subtle, probabilistic dimension of privacy loss. \edDP\ is sometimes also called \textit{approximate differential privacy}~\cite{meiser_approximate_2018}.

\begin{samepage}
\begin{mydef}[(\ep, $\delta$)-Differential Privacy]\label{def:edDP}
A randomized algorithm $f'$ is (\ep, $\delta$)-differentially private if for all data sets $D_1$ and $D_2$ differing on at most one element, and all $\mathcal{S}$ $\subseteq$ Range($f'$),
\[
\text{Pr}[f'(D_1) \in \mathcal{S}]\leq e^{\mathepsilon} \times \text{Pr}[f'(D_2) \in \mathcal{S}] + \delta
\]
\end{mydef}
\end{samepage}

Theoretically, in \eDP\ each output is \textit{nearly} equally likely and hold for \textit{any} run of algorithm $f'$, whereas \edDP\ for \textit{each pair} of \dataset s $(D_1,D_2)$ in extremely unlikely cases, will make some answer much less or much more likely to be released when the algorithm is run on $D_1$ as opposed to $D_2$~\cite{dwork_algorithmic_2014}. Still, \edDP\ ensures that the absolute value of the privacy loss is bounded by \ep\ with probability at least 1-$\delta$~\cite{dwork_algorithmic_2014}. That is, the probability of gaining significant information about one individual, even when possessing all other information in the \dataset, is at most $\delta$.

To satisfy differential privacy, a randomized algorithm perturbs the query answers to obfuscate the impact caused by differing one element in the \dataset. Such perturbation can for example be introduced by adding a randomly chosen number to a numerical answer. Essentially, the maximum difference \textit{any} possible record in the \dataset can cause dictates the magnitude of noise needed to satisfy differential privacy. This difference is referred to as the algorithm's $L_1$ sensitivity, which we define in \Cref{def:sensitivity}, based on \citet{dwork_calibrating_2006}.

\begin{samepage}
\begin{mydef}[$L_1$ Sensitivity~]\label{def:sensitivity}
The $L_1$ sensitivity of a function $f$ : $D^n \to \mathbb{R}^d$
is the smallest number $\Delta f$ such that for all $D_1$, $D_2 \in D^n$  which differ in a single entry,
\[
{\|f (D_1) - f (D_2)\|}_1 \leq \Delta f
\]
\end{mydef}
\end{samepage}

Since differential privacy is a property of the algorithm, as opposed to data, there exists many implementations of differentially private algorithms. Thus, we will not summarize all algorithms, but instead introduce two early algorithms that are common building blocks, namely: the Laplace mechanism~\cite{dwork_calibrating_2006} and the Exponential mechanism~\cite{mcsherry_mechanism_2007}.

We define the Laplace mechanism in \Cref{def:laplace}, based on the definition given by \citet{dwork_differential_2008}. The Laplace mechanism adds numerical noise, and the probability density function is centered around zero, meaning that noise with higher probability (than any other specific value) will be zero. 

\begin{mydef}[Laplace mechanism]\label{def:laplace}
For a query $f$ on \dataset\ $D$, the differentially private version, $f'$, adds Laplace noise to $f$ proportional to the sensitivity of $f$:
\[
f'(D) = f(D) + Lap(\Delta f /\mathepsilon)
\]
\end{mydef}

Furthermore, we define the Exponential mechanism (EM) in \Cref{def:em} based on the definition given by \citet{mcsherry_mechanism_2007}. The intuition behind EM is that the probability of not perturbing the answer is slightly higher than perturbing the answer. EM is particularly useful when Laplace does not make sense, for example when queries return categorical answers such as strings, but can also be used for numerical answers. The reason EM is so flexible is that the utility function can be replaced to score closeness to suit the given domain.

\begin{mydef}[Exponential mechanism (EM)]\label{def:em}
Given a utility function $u : (D \times R) \to R$, and a \dataset\ $D$, we define the differentially private version, $u'$:
\[
u'(D,u) = \bigg\{\text{return }r,\text{where }r \text{ ranges over }R, \text{ with probability }\propto exp\frac{\mathepsilon u(D,r)}{2\Delta u} \bigg\}
\]
\end{mydef}

% "Neighboring" databases, what does it mean?
The semantic interpretation of the privacy guarantee of differential privacy rests on the definition of what it means for a pair of \dataset s to be neighbors. In the literature, the following two variations of neighbors are considered when defining differential privacy: unbounded and bounded.

%In Definition~\ref{def:eDP}, the concept of neighboring \dataset s is intentionally kept general, and therefore includes both of the following cases:
%Let $D_{I}$ be a \dataset\ containing $I$ individuals,
\begin{mydef}[]Let $D_{1}$ and $D_{2}$ be two data sets where $D_1$ can be attained by adding or removing a single record in $D_2$. With this notion of neighbors, we say that we have \textit{unbounded} differential privacy.
\end{mydef}
\begin{mydef}[] Let $D_{1}$ and $D_{2}$ be two data sets where $D_1$ can be attained by changing a single record in $D_2$. With this notion of neighbors, we say that we have \textit{bounded} differential privacy.
\end{mydef}
% \begin{mydef}[Unbounded differential privacy] Dataset $D_{J} = D_{I\pm x}$ for any x in the population then $D_{J}$ is a neighboring \dataset\ of $D_{I}$ 
% \end{mydef}
% \begin{mydef}[Bounded differential privacy] Dataset $\forall\ x\ \in I, D_{J}$  is said to be  $D_{I}$ iff, $D_{I}(i)$ = $D_{J}(j)$ for any one element $i\ \in I$
% \end{mydef}

Distinguishing between the definition of neighboring \dataset s is important, because it affects the global sensitivity of a function. The sizes of the neighboring \dataset s are fixed in the bounded differential privacy definition whereas, there is no size restriction in the unbounded case.

%For example, the sensitivity of a histogram function in unbounded differential privacy is 1, whereas in the bounded setting, the sensitivity is 2. The reason for this difference is that in the bounded setting, the size of the \dataset\ is fixed, whereas in the unbounded case the size can change arbitrarily.

In the case of graph \dataset s, a pair of graphs differ by their number of edges, or number of nodes. Therefore, there exists two variant definitions in literature~\cite{Hay_degree_distribution_2009} that formalize what it means for a pair of graphs to be neighbors. Nevertheless, these graph neighborhood definitions are defined only in the context of unbounded differential privacy.

\begin{mydef}[Node differential privacy~\cite{Hay_degree_distribution_2009}]\label{case:nodeDP}
Graphs $G=(V,E)$ and $G^{'} = (V^{'},E^{'})$ are \emph{node-neighbors} if:
\begin{align*}
    V^{'} &= V - v,\\
    E^{'} &= E - \{(v_1,v_2)\mid v_1\ =\ v \vee v_2\ =\ v\},
\end{align*}
for some node $v\ \in\ V.$
\end{mydef}
   
\begin{mydef}[Edge differential privacy~\cite{Hay_degree_distribution_2009}] \label{case:edgeDP}
Graphs $G=(V,E)$ and $G^{'}=(V^{'},E^{'})$ are \emph{edge-neighbors} if:
\begin{align*}
   V\ &=\ V^{'}, \\
   E^{'} &= E - \{e\},\
\end{align*}
for some edge $e \in E$.
\end{mydef}

%Epsilon, trade-off w/ accuracy
In certain settings, \ep\ grows too fast to guarantee a meaningful privacy protection. To cater to different applications, in particular in settings where data is gathered dynamically, different \textit{privacy levels} have been introduced that essentially further changes the notion of neighboring \dataset s by defining neighbors for data streams. These privacy levels are, user level privacy~\cite{dwork_pan-private_2010}, event level privacy~\cite{dwork_differential_2010}, and \wevent~\cite{kellaris_differentially_2014}.

\begin{mydef}
We say that a differentially private query gives \textit{user level privacy} (pure differential privacy), when all occurrences of records produced by one user is either present or absent.
\end{mydef}

Essentially, for user level privacy, all records connected to one individual user shares a joint privacy budget.

\begin{mydef}
We say that a differentially private query gives \textit{event level privacy}, when all occurrences of records produced by one group of events, where the group size is one or larger, is either present or absent.
\end{mydef}

With event level privacy, each data point used in the query can be considered independent and thus have their own budget.

\begin{mydef}
We say that a differentially private query gives $\mathit{w}$\textit{-event level privacy}, when a set of $\mathit{w}$ occurrences of records produced by some group of events, where the group size is one or larger, is either present or absent. When $\mathit{w}=1$, \wevent\ and \event\ are the same.
\end{mydef}

For $\mathit{w}$-event level privacy, $\mathit{w}$ events share a joint privacy budget.
\section{Method}\label{sec:method}
\begin{figure}[htbp]
    \centering
    \includegraphics[width=\linewidth]{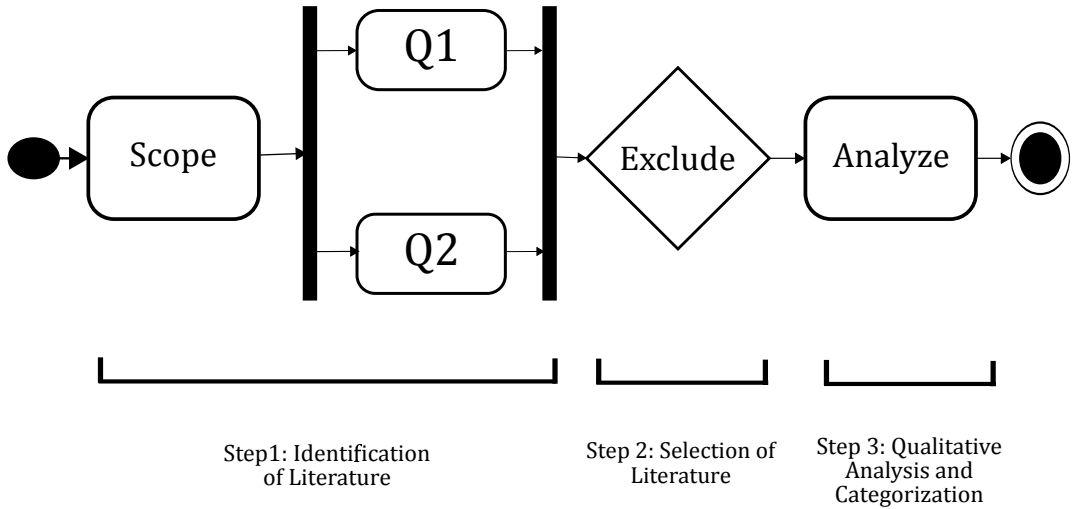}
    \caption{Workflow of processes followed in our SLR.}
    \label{fig:process}
\end{figure}

We used Systematic Literature Review (SLR)~\cite{kitchenham_procedures_2004} to synthesize the state-of-the-art accuracy improving techniques for differentially private histogram and synthetic data publication. SLR provides methodological rigor to literature selection and synthetization as well as to the conclusion drawn as a result of the synthetization. Consequently, SLR generally consists of three steps: identification of literature, selection of literature for review and analysis of the state-of-the-art. \Cref{fig:process} shows the high-level view of the processes followed in our SLR. In the subsequent subsections, we describe each processes in detail.

\subsection{Identification of Literature}
A thorough and unbiased search for literature is the essence of a SLR. In this SLR, we used a scholarly search engine, Microsoft Academic~\cite{sinha_overview_2015, microsoft_microsoft_nodate}, primarily for two reasons. First, for its semantic search functionality and second, for its coverage. Semantic search leverages \textit{entities} such as field of study, authors, journals, institutions, etc., associated with the papers. Consequently, there is no need to construct search strings with more keywords and synonyms, rather, a natural language query can be constructed with the help of search suggestions for relevant entities. We used two queries, one focusing on histograms and the other on synthetic data. The queries are as follows, with entities recognized by Microsoft Academic in bold text:

\begin{itemize}
    \item[\textbf{Q1}:] \textbf{Papers} about \textbf{differential privacy} and \textbf{histograms}
    \item[\textbf{Q2}:] \textbf{Papers} about \textbf{differential privacy} and \textbf{synthetic data}
\end{itemize}

The search was performed on June 10 2019, and yielded 159 hits in total. 78 hits for \textbf{Q1} and 81 hits for \textbf{Q2}, which are examined for relevance in the next step of the SLR process.

\subsection{Selection of Literature}
We constructed and followed a set of exclusion criteria \Cref{tab:exclusion} in order to select the relevant literature that provides insights to our research aim. To reflect that we specifically wanted to focus on tangible, experimentally tested algorithms, we constructed the criteria to exclude papers that contribute pure theoretical knowledge.

To select papers, we examined the title and abstract of each paper against the exclusion criteria. When the abstract matches any one of the criteria, the paper is excluded, otherwise the paper is included. When it was unclear from the abstract that a contribution is empirical or pure theory, we looked through the body of the paper to make our decision. For the full list of excluded papers along with the reason for exclusion, see \Cref{app:appendix}.

\begin{table}[htbp]
    \centering\rowcolors{1}{}{gray!10}
    \begin{tabular}{p{\linewidth}}
    \textbf{Exclude if the paper is...}\\
   \textbf{1)} not concerning differential privacy, not concerning accuracy improvement, and not concerning histogram or synthetic data.\\
    \textbf{2)} employing workflow actions, pre-processing/post-processing/algorithmic tricks but not solely to improve accuracy of histogram or synthetic data.\\
    \textbf{3)} a trivial improvement to histogram or synthetic data accuracy through relaxations of differential privacy or adversarial models.\\
    \textbf{4)} concerning local sensitivity as opposed to global sensitivity.\\
    \textbf{5)} not releasing histogram/synthetic data.\\
    \textbf{6)} pure theory, without empirical results.\\
    \textbf{7)} about a patented entity.\\
    \textbf{8)} a preprint or otherwise unpublished work.\\
    \textbf{9)} not peer reviewed such as PhD thesis/master thesis/demo paper/poster/extended abstract.\\
    \textbf{10)} not written in English.
    \end{tabular}
    \caption{List of exclusion criteria followed in our SLR.}
    \label{tab:exclusion}
\end{table}

In the end, a total of 35 (after removing duplicates) papers were selected for the qualitative analysis.

% what is the result of this step.

%The list of exclusion criteria, was manually checked against the abstract of each paper. When a paper matches any one of the criteria, it is excluded, otherwise the paper is included. For the full list of excluded papers and why they were excluded, see \Cref{app:appendix}. %read the abstract, if it doesn't mention accuracy/improvement/synonym at all, exclude, and if it's unclear if there's empirical results/theory, look through the corpus of the paper

\subsection{Qualitative analysis and Categorization}
The most common framework found in the literature to analyse and understand a domain of interest, is classification schemes~\cite{nickerson_method_2013}. It concerns the grouping of objects with similar characteristics in a domain. Our aim is to synthesize; i) on the one hand, trends and relationships among each papers and ii) on the other hand, conceptual understanding of the privacy/accuracy \tradeoff\ in the differentially private histogram and synthetic data research. Therefore, from each paper we extracted distinct characteristics of the algorithms, evaluation details of the algorithms as well as design principles such as aim of the solution and motivation for using a particular technique. These characteristics are inductively analyzed for commonality, which follows, though not rigorously, the empirical-to-conceptual approach to taxonomy development defined by \citet{nickerson_method_2013}. The categorization that resulted from the qualitative analysis are presented in \Cref{sec:analysis}.

\section{Overview of Papers}\label{sec:paper-summaries}

After analyzing the 35 included papers, 27 papers~\citep{hay_boosting_2010,
ding_differentially_2011,
xiao_differential_2011,
acs_differentially_2012,
xu_differentially_2013,
li_differentially_2014,
lu_generating_2014,
park_pegs_2014,
xiao_dpcube_2014,
zhang_privbayes_2014,
zhang_towards_2014,
chen_private_2015,
lee_maximum_2015,
li_differentially_2015,
day_publishing_2016,
wang_private_2016,
zhang_privtree_2016,
benkhelif_co-clustering_2017,
doudalis_sortaki_2017,
kotsogiannis_pythia_2017,
wang_differentially_2017,
xu_dppro_2017,
ding_privacy-preserving_2018,
gao_dynamic_2018,
ghane_publishing_2018,
li_ihp_2019,
nie_utility-optimized_2019} where found to be relevant. All included papers and their corresponding algorithms are listed in the ledger in \Cref{tab:included}. We illustrate the chronological publishing order of the algorithms, but note that within each year, the algorithms are sorted on the first author's last name, and not necessarily order of publication. 

\begin{table}[h!tb]
\centering
\begin{tabular}{@{\,}r <{\hskip 2pt} !{\foo} >{\raggedright\arraybackslash}l}
2010 & \begin{tabularx}{9cm}{Xr}\rowcolor{gray!10}
    \boost & \citet{hay_boosting_2010} 
    \end{tabularx}\\
\arrayrulecolor{blue1}\hline
2011 & \rowcolors{1}{}{gray!10}\begin{tabularx}{9cm}{Xr}
     \pmost, \bmax  & \citet{ding_differentially_2011} \\
     \privelet, \priveletplus, \priveletstar & \citet{xiao_differential_2010,xiao_differential_2011} 
\end{tabularx}\\
\arrayrulecolor{blue1}\hline
2012 & \rowcolors{1}{}{gray!10}\begin{tabularx}{9cm}{Xr}
\efpa, \php & \citet{acs_differentially_2012}
\end{tabularx}\\
\arrayrulecolor{blue1}\hline
2013 & \begin{tabularx}{9cm}{Xr}\rowcolor{gray!10}
\noisefirst, \structurefirst & \citet{xu_differentially_2012,xu_differentially_2013} 
\end{tabularx}\\
\arrayrulecolor{blue1}\hline
2014 & \rowcolors{1}{}{gray!10}\begin{tabularx}{9cm}{Xr}
 \dpcopula & \citet{li_differentially_2014} \\
\citm  & \citet{lu_generating_2014} \\
\pegs, \pegsrs  & \citet{park_perturbed_2013}, \citet{park_pegs_2014} \\
\dpcube  & \citet{xiao_differentially_2010, xiao_dpcube_2012-1,xiao_dpcube_2014} \\
\privbayes  & \citet{zhang_privbayes_2014} \\
\ahp & \citet{zhang_towards_2014} 
\end{tabularx}\\
\arrayrulecolor{blue1}\hline
2015 & \rowcolors{1}{}{gray!10}\begin{tabularx}{9cm}{Xr}
\rg & \citet{chen_private_2015} \\
\admm & \citet{lee_maximum_2015} \\
\dsat, \dsft & \citet{li_differentially_2015} 
\end{tabularx}\\
\arrayrulecolor{blue1}\hline
2016 & \begin{tabularx}{9cm}{Xr}
\rowcolor{gray!10}
\thetaomegahistogram, \thetacumhisto & \citet{day_publishing_2016} \\
\rowcolor{white}
\bpm & \citet{wang_private_2016} \\
\rowcolor{gray!10}
\privtree & \citet{zhang_privtree_2016} 
\end{tabularx}\\
\arrayrulecolor{blue1}\hline
2017 & \rowcolors{1}{}{gray!10}\begin{tabularx}{9cm}{Xr}
\dpcocgen & \citet{benkhelif_co-clustering_2017} \\
\sortaki & \citet{doudalis_sortaki_2017} \\
\pythia, \delphi & \citet{kotsogiannis_pythia_2017} \\
\tru, \minalg, \opt & \citet{wang_differentially_2017} \\
\dppro & \citet{xu_dppro_2017} 
\end{tabularx}\\
\arrayrulecolor{blue1}\hline
2018 & \begin{tabularx}{9cm}{Xr}
\rowcolor{gray!10}
\tlambda & \citet{ding_privacy-preserving_2018}  \\
\rowcolor{white}
\gga & \citet{gao_dynamic_2018} \\
\rowcolor{gray!10}
\prish & \citet{ghane_publishing_2018} \\
\end{tabularx}\\
\arrayrulecolor{blue1}\hline
2019 & \rowcolors{1}{}{gray!10}\begin{tabularx}{9cm}{Xr}
\ihp, \mihp & \citet{li_improving_2016,li_ihp_2019} \\
\rcf & \citet{nie_utility-optimized_2019} 
\end{tabularx}
\end{tabular}
\caption{Chronological ledger for the papers. Note that the abbreviation 'ADMM' is due to \citet{boyd_distributed_2011}, whereas \citet{lee_maximum_2015}'s work is an extension that uses the same abbreviation}\label{tab:included}
\end{table}

Beware that some algorithms, for example \noisefirst, \structurefirst, have appeared in publications twice, first in a conference paper and then in an extended journal version. When a paper has two versions, we will refer to the latest version in our comparisons, but we include all references in the paper ledger for completeness. Furthermore, eight papers were excluded based on our qualitative analysis. Each decision is motivated in \Cref{sec:incomparable}, and those eight papers hence do not appear in the paper ledger.

\begin{table}[htbp]
\centering
\rowcolors{3}{}{gray!10}\resizebox{1\textwidth}{!}{%
\begin{tabularx}{\textwidth}{p{4mm}p{6mm}p{10mm}p{12mm}p{6mm}p{7mm}p{15mm}p{18mm}X}
\toprule
    \textbf{Ref.}& \textbf{Def.} & \textbf{Lvl.} & \textbf{Rel.} & \textbf{Dim.} & \textbf{In.} & \textbf{Mech.} & \textbf{Metric} & \textbf{Out.}  \\ \midrule
\cite{hay_boosting_2010} & \ep & ? & \unboundedsymbol & \textsc{1d} & \staticsymbol & \textsc{Lap} & \textsc{mae} & \textsc{Histogram}\\
\cite{ding_differentially_2011} & \ep & ? & \unboundedsymbol & * & \staticsymbol, \correlatedsymbol & \textsc{Lap} & \textsc{mae} & \textsc{Cuboids} \\
\cite{xiao_differential_2011} & \ep & ? & \boundedsymbol &  \textsc{1d}, * & \staticsymbol & \textsc{Lap} & \textsc{mae}, \textsc{mpe} &  \textsc{Range count queries} \\ 
\cite{acs_differentially_2012} & \ep & ? & \unboundedsymbol & \textsc{1d} & \staticsymbol, \correlatedsymbol & \textsc{Lap}, \textsc{em} & \textsc{kl}, \textsc{mse} & \textsc{Histogram} \\
\cite{xu_differentially_2013} & \ep & ? & \unboundedsymbol & \textsc{1d} & \staticsymbol & \textsc{Lap}, \textsc{em} & \textsc{mae}, \textsc{mse} & \textsc{Histogram} \\
\cite{li_differentially_2014} & \ep & ? & \unboundedsymbol & *, \sparsesymbol & \staticsymbol & \textsc{Lap} & \textsc{mae}, \textsc{mpe} & \textsc{Synthetic data}\\
\cite{lu_generating_2014} & (\ep,$\delta$) & \textsc{entity} & \unboundedsymbol & * & \staticsymbol, \correlatedsymbol & \parbox{4em}{\textsc{mm},\\\textsc{agnostic}} & \textsc{mpe} & \textsc{Model} \\
\cite{park_pegs_2014} & \ep & ? & \unboundedsymbol & * & \staticsymbol & \textsc{Dirichlet prior} & \parbox{2em}{\textsc{Rank}\\\textsc{corr.}} & \textsc{Model} \\
\cite{xiao_dpcube_2014} & \ep & ? & \unboundedsymbol & * & \staticsymbol & \textsc{Lap} & \textsc{mae} & \textsc{Histogram} \\
\cite{zhang_privbayes_2014} & \ep & ? & \boundedsymbol &  *, \sparsesymbol & \staticsymbol & \textsc{Lap}, \textsc{em} & \textsc{avd}, \textsc{Miss} & \textsc{Synthetic data} \\
\cite{zhang_towards_2014} & \ep & ? & \unboundedsymbol & \textsc{1d} & \staticsymbol & \textsc{Lap} & \textsc{kl}, \textsc{mse} & \textsc{Histogram} \\
\cite{chen_private_2015} & \ep & \textsc{Event} & \unboundedsymbol & \textsc{1d} &  \dynamicsymbol, \correlatedsymbol & \textsc{Lap} & \textsc{mse} & \textsc{Histogram} \\
\cite{lee_maximum_2015} & \ep & ? & \unboundedsymbol & * & \staticsymbol & \textsc{Lap}, \textsc{mm} & \textsc{mse} & \textsc{Contingency table}, \mbox{\textsc{histogram}} \\
\cite{li_differentially_2015} & \ep & \parbox{4em}{ \textsc{user},\\ $\mathit{w}$-\textsl{event}} & \unboundedsymbol &\textsc{1d} & \dynamicsymbol, \correlatedsymbol & \textsc{Lap} & \textsc{mae}, \textsc{mpe} & \textsc{Histogram} \\
\cite{day_publishing_2016} & \ep & ? & \parbox{4em}{ \textsc{node}\\ \textsc{privacy}} & * & \staticsymbol & \textsc{em} & \textsc{ks}, $\ell1$ & \textsc{Histogram} \\ 
\cite{wang_private_2016} & \ep & ? & \boundedsymbol & \textsc{1d} & \staticsymbol & \textsc{rr} & \textsc{nwse} & \textsc{Histogram} \\
\cite{zhang_privtree_2016} & \ep & ? & \unboundedsymbol & * &  \staticsymbol, \correlatedsymbol & \textsc{Lap} & \textsc{mpe} & \textsc{Quadtree} \\
\cite{benkhelif_co-clustering_2017} & \ep & ? & \unboundedsymbol & * &  \staticsymbol, \sparsesymbol & \textsc{Lap} & \textsc{Hellinger} & \textsc{Partitioning} \\
\cite{doudalis_sortaki_2017} & \ep & ? & \unboundedsymbol & \textsc{1d} & \staticsymbol & \textsc{Lap} & \textsc{saq} & \textsc{Histogram}\\
\cite{kotsogiannis_pythia_2017} & \ep & ? & \unboundedsymbol &  \textsc{1d}, * & \staticsymbol & \parbox{4em}{ \textsc{Lap},\\ \textsc{agnostic}} & $\ell2$, \textsc{regret} & \textit{N/A}\\
\cite{wang_differentially_2017} & \ep & ? & \unboundedsymbol & \textsc{1d} &  \staticsymbol, \correlatedsymbol & \textsc{Lap} & \textsc{mse} & \textsc{Histogram}\\
\cite{xu_dppro_2017} & (\ep,$\delta$) & ? & \boundedsymbol & * & \staticsymbol & \parbox{4em}{ \textsc{Gaussian},\\ \textsc{mm}} & \textsc{Miss}, \textsc{mse} & \textsc{Matrix} \\
\cite{ding_privacy-preserving_2018} & \ep & ? & \parbox{4em}{ \textsc{node}\\ \textsc{privacy}} & * & \staticsymbol & \textsc{Lap} &  \textsc{ks}, $\ell1$ & \textsc{Histogram} \\
\cite{gao_dynamic_2018} & \ep & ? & \unboundedsymbol & \textsc{1d} & \dynamicsymbol & \textsc{Lap} & \textsc{mae} & \textsc{Histogram} \\
\cite{ghane_publishing_2018} & \ep & ? & \unboundedsymbol & *, \sparsesymbol & \staticsymbol, \correlatedsymbol & \textsc{mwem} & \textsc{kl}, $\ell1$ & \textsc{Histogram} \\
\cite{li_ihp_2019} & \ep & ? & \unboundedsymbol &  \textsc{1d},* & \staticsymbol, \sparsesymbol & \textsc{Lap}, \textsc{em} & \textsc{kl}, \textsc{mse} & \textsc{Histogram} \\
\cite{nie_utility-optimized_2019} &\ep & ? & \boundedsymbol & \textsc{1d} & \staticsymbol & \textsc{rr} & \textsc{mse} & \textsc{Histogram}\\
\end{tabularx}%
}
    \caption{Mapping between papers to corresponding differential privacy definition, privacy level, neighbor relationship, dimension of data, input data, use of mechanism, error metric and output data. Abbreviations and the corresponding symbols are explained in a separate table.}
    \label{tab:facts}
\end{table}

\begin{table}[htb]
    \centering
    \begin{tabularx}{\linewidth}{X|c|l}
    \toprule
        & \textbf{Key} & \textbf{Meaning} \\
        \midrule
        {\multirow{4}{*}{\textbf{Data}}} & \correlatedsymbol& Correlated\\
        & \dynamicsymbol & Dynamic \\
        &\sparsesymbol & Sparse \\
        & \staticsymbol& Static \\
        \midrule
         {\multirow{2}{*}{\textbf{Dimension}}} & * & Multi\\
        & \textsc{1d} & Single \\
        \midrule
        {\multirow{4}{*}{\textbf{Mechanism}}}&\textsc{em}&Exponential mechanism\\
         & \textsc{Lap} & Laplace mechanism \\
         &\textsc{mm} & Matrix mechanism \\
        &\textsc{rr} &Randomized response \\
        \midrule
         {\multirow{10}{*}{\textbf{Metric}}}&\textsc{avd} & Average Variation Distance\\
        &\textsc{ks} & Kolmogorov-Smirnov distance \\
        &\textsc{kl} & Kullback-Leibler divergence \\
        & $\ell1$ & L1 distance\\
        & $\ell2$ & L2 distance\\
        &\textsc{mae} & Mean absolute error \\
        & \textsc{Miss} & Misclassification rate\\
        &\textsc{mpe} & Mean percentage error\\
        &\textsc{mse} & Mean squared error\\
        &\textsc{nwse} & Normalized weighted square error\\
        &\textsc{saq} & Scaled average per query\\
        \midrule
        {\multirow{2}{*}{\textbf{Relation}}} & \boundedsymbol & Bounded \\
        & \unboundedsymbol& Unbounded\\
        \bottomrule
    \end{tabularx}
    \caption{Meaning of symbols and abbreviations }
    \label{tab:facts-abbreviations}
\end{table}

Furthermore, in \Cref{tab:facts,tab:facts-abbreviations}, we present objective parameters regarding the settings around the algorithms in each paper, for example the characteristics of the input data they operate on, and the metric used to measure errors. Our intention is that this table will allow for further understanding of which algorithms are applicable given a certain setting when one searches for an appropriate algorithm, but also to understand which algorithms are directly comparable in the scope of this SLR.

Note that the privacy level (user, event or $\mathit{w}$-event) was not explicitly stated in most papers, in which case we have attributed the privacy level as '?'. A '?' privacy level does not imply that the algorithm does not have a particular privacy level goal, but rather, that the authors did not explicitly describe what level they are aiming for. With this notice, we want to warn the reader to be cautious when comparing the experimental accuracy of two algorithms unless they in fact assume the same privacy level. For example, comparing the same algorithm but with either user level or event level privacy would make the event level privacy version appear to be better, whereas in reality it trivially achieves better accuracy through relaxed privacy guarantees.

In general, user level privacy tends to be the base case, as this is the level assumed in \textit{pure} differential privacy~\cite{dwork_pan-private_2010}, but to avoid making incorrect assumptions, we chose to use the '?' label when a paper does not explicitly state their privacy level.

\begin{table}[h!tb]
    \centering
    \begin{tabularx}{\linewidth}{r>{\columncolor{gray!10}}cl}
    \toprule
         \textbf{Histogram} & \textbf{Hybrid} & \textbf{Synthetic Data}  \\\midrule
         \citet{hay_boosting_2010}& 
         \citet{lu_generating_2014} & 
         \citet{li_differentially_2014}\\
         \citet{xiao_differential_2011}&
         \citet{ding_differentially_2011}&
          \citet{park_pegs_2014} \\
         \citet{acs_differentially_2012}&
          \citet{xiao_dpcube_2014}&
          \citet{zhang_privbayes_2014}\\
         \citet{xu_differentially_2013}&
         \citet{lee_maximum_2015}&
         \citet{xu_dppro_2017} \\
          \citet{zhang_towards_2014}&
          \citet{zhang_privtree_2016} &
          \\
         \citet{chen_private_2015}&
         \citet{benkhelif_co-clustering_2017} &
         \\
        \citet{li_differentially_2015} & 
        \citet{kotsogiannis_pythia_2017} &
        \\
        \citet{day_publishing_2016} &
        \citet{wang_differentially_2017} & \\
         \citet{wang_private_2016} & 
        \citet{li_ihp_2019} & \\
         \citet{doudalis_sortaki_2017}& 
         & \\
         \citet{ding_privacy-preserving_2018}& & \\
        \citet{gao_dynamic_2018} & & \\
        \citet{ghane_publishing_2018} & & \\
        \citet{nie_utility-optimized_2019} & & \\
        \bottomrule
    \end{tabularx}
    \caption{The papers grouped by their type of output, where hybrid internally uses histogram structures where synthetic data is sampled from}
    \label{tab:paper-overview}
\end{table}

Given that our two queries were designed to capture algorithms that either output synthetic data or histogram, we examine the similarity between the strategies used in each algorithm. To this end, we manually represent the similarity between the algorithms' strategies based on their output in \Cref{tab:paper-overview}. We distinguish between the two kinds of outputs by their different goals: for histograms, the goal is to release \textit{one optimal histogram} for a given query, whereas for synthetic data the goal is to release a \dataset\ that is optimized for some \textit{given set of queries}. Some algorithms use similar approaches to the algorithms from the other query; and therefore we label them as hybrid. An example of a hybrid paper is \citet{li_ihp_2019}, since they both deal with one-dimensional histograms (\ihp), and then re-use that strategy when producing multi-dimensional histograms (\mihp) that resembles the outputs of synthetic data papers.

\section{Analysis}\label{sec:analysis}

We present our qualitative analysis on 27 included papers from two different perspectives in the light of research in differential privacy histogram and synthetic data. First, from a evolutionary perspective for identifying trends and to position each contribution in the history of its research (\Cref{sec:positioning}). Second, from a conceptual perspective for understanding the \tradeoff\ challenge in the privacy and utility relationship (\Cref{sec:dimensions}).

\subsection{Positioning}\label{sec:positioning}
In order to provide context, we studied \textit{where} the algorithms originated from, and how they are connected to each other. To also understand \textit{when} to use each algorithm, and which ones are comparable in the sense that they can be used for the same kind of analysis, we also investigate which algorithms are compared experimentally in the papers. 

\begin{figure}[htb]
    \centering
    \includegraphics[width=\linewidth, height=8.5cm]{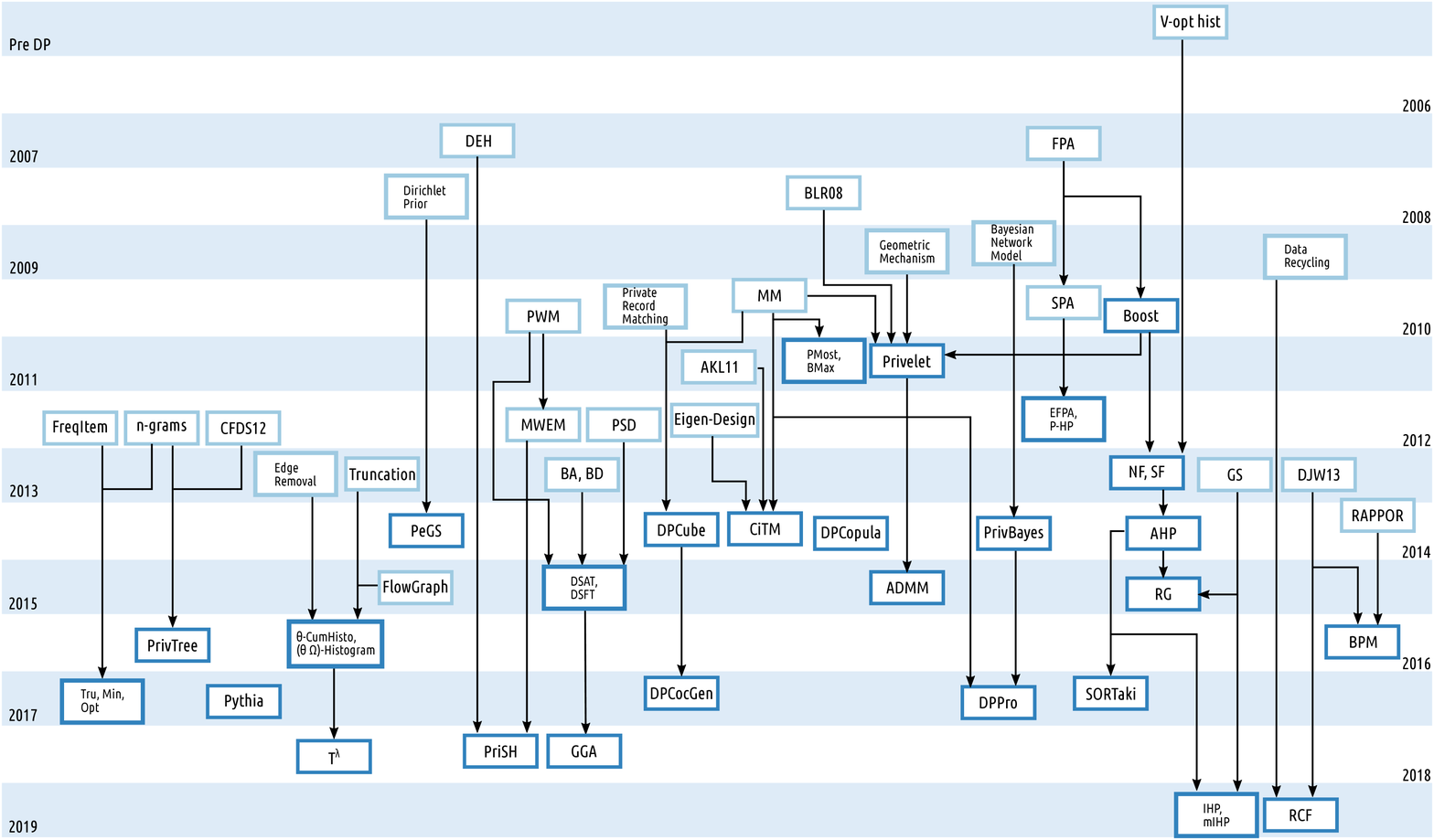}
    \caption{The family tree of algorithms. Light blue indicate papers not covered by the SLR, and the darker blue represents included papers.}
    \label{fig:family-tree}
\end{figure}

First, we explored and mapped out the relationships between the included algorithms. To further paint the picture of the landscape of algorithms, we analyzed the related work sections to find external work connected to the papers included in our SLR. We present our findings as a family tree of algorithms in \Cref{fig:family-tree}, which addresses from where they came.

\begin{table}[htb]
\centering\rowcolors{1}{}{gray!10}
\begin{tabularx}{\linewidth}{rl}\toprule
\textbf{Label} & \textbf{Author} \\ \midrule
AKL11 & \citet{arasu_data_2011}\\
Bayesian Network Model & \citet{koller_probabilistic_2009}\\
BLR08 & \citet{blum_learning_2008} \\
Budget Absorption (BA), Budget Distribution (BD) & \citet{kellaris_differentially_2014} \\
CFDS12 & \citet{chen_differentially_2012-1}\\
Data Recycling & \citet{xiao_optimal_2009}\\
Dirichlet Prior & \citet{machanavajjhala_privacy_2008}\\
Distributed Euler Histograms (DEH) & \citet{xie_distributed_2007}\\
DJW13 & \citet{duchi_local_2013}\\
Edge Removal & \citet{blocki_differentially_2013} \\
Eigen-Design & \citet{li_adaptive_2012}\\
FlowGraph & \citet{raskhodnikova_efficient_2015}\\
Fourier Perturbation Algorithm (FPA) & \citet{barak_privacy_2007}\\
FreqItem & \citet{zeng_differentially_2012}\\
Geometric Mechanism & \citet{ghosh_universally_2012}\\
Grouping and Smoothing (GS) & \citet{kellaris_practical_2013}\\
Matrix Mechanism (MM) & \citet{li_optimizing_2010}\\
MWEM & \citet{hardt_simple_2012}\\
n-grams & \citet{chen_differentially_2012}\\
Private Multiplicative Weights (PMW) & \citet{hardt_multiplicative_2010}\\
Private Record Matching & \citet{inan_private_2010}\\
Private Spatial Decompositions (PSD) & \citet{cormode_differentially_2012} \\
RAPPOR & \citet{erlingsson_rappor_2014}\\
Sampling Perturbation Algorithm (SPA) & \citet{rastogi_differentially_2010}\\
Truncation & \citet{kasiviswanathan_analyzing_2013}\\
V-opt hist & \citet{jagadish_optimal_1998}\\
\end{tabularx}
\caption{Ledger for papers outside of the SLR}\label{tab:external}
\end{table}

Since our exploration of each algorithms' origin discovered papers outside of the SLR's queries, we also provide a ledger for (\Cref{tab:external}) \textit{external} papers. When the authors had not designated a name for their algorithms, we use the abbreviation of the first letter of all author's last name and the publication year instead. Note that we have not recursively investigated the external papers' origin, so external papers are not fully connected in the family tree.

From the family tree, we notice that there are several different lines of research present. One frequently followed line of research is that started by \citet{xu_differentially_2013}, \noisefirst, \structurefirst, which addresses the issue of finding an appropriate histogram structure (i.e. bin sizes) by creating a differentially private version of a v-optimal histogram. Essentially, EM is used to determine the histogram structure, and then the Laplace mechanism is used to release the bin counts. The idea by \citet{xu_differentially_2013} is followed by \ahp, \rg, \sortaki, \ihp\ and \mihp.

The matrix mechanism (MM) is a building block that is used in \pmost, \bmax, \citm\ and \dppro. Apart from using the same release mechanism, they do not share many similarities as also becomes apparent when comparing their experimental evaluation.

Only \pythia\ and \dpcopula\ appears as orphaned nodes in the family tree. \pythia\ is special in the sense that it is not a standalone algorithm, but rather provides a differentially private way of choosing the 'best' algorithm for a given \dataset. \dpcopula\ has a mathematical background in copula functions, which are functions that describe the dependence between multivariate variables. This approach of using copula functions is not encountered in any of the other papers.

To further put the algorithms into perspective, we explored which algorithms were used in their experimental comparisons. The comprehensive matrix of which algorithms are experimentally compared to each other in \Cref{tab:comparison-matrix}. This complements the fact table (\Cref{tab:facts}) in addressing the question of when to use an algorithm, as algorithms that are compared experimentally can be used interchangeably for the same analysis. E.g, when \noisefirst\ is used, it can be swapped with for example \ihp. 

\begin{table}[H]
    \centering\rowcolors{1}{}{gray!10}
    \begin{tabularx}{\linewidth}{XXl}
    \toprule
        \textbf{Algorithm} & \textbf{Internal Comparison} & \textbf{External Comparison} \\
        \midrule
\boost &- &-\\
\pmost, \bmax & - & -\\ 
\privelet, \priveletplus, \priveletstar & -& -\\
\efpa, \php & \boost, \privelet,\noisefirst, \structurefirst & SPA~\cite{rastogi_differentially_2010}, MWEM~\cite{hardt_simple_2012} \\
\noisefirst, \structurefirst & \boost, \privelet & - \\
\dpcopula & \priveletplus, \php& FP~\cite{cormode_differentially_2012-1}, PSD~\cite{cormode_differentially_2012}\\
\citm & -& -\\
\pegs, \pegsrs &- &-\\
\dpcube & \boost & \parbox{14em}{Private Interactive
ID3~\cite{friedman_data_2010},\\ Private Record Matching~\cite{inan_private_2010}}\\
\privbayes & - &\parbox{10em}{FPA\cite{barak_privacy_2007},\\ PrivGene~\cite{zhang_privgene_2013},\\ ERM~\cite{chaudhuri_differentially_2011}}\\
\ahp & \noisefirst, \structurefirst, \php & GS~\cite{kellaris_practical_2013}\\
\rg & - & BA~\cite{kellaris_differentially_2014}, FAST~\cite{fan_adaptive_2014}\\
\admm & \boost, \efpa, \php, \privelet & LMM~\cite{li_optimizing_2010}, RM~\cite{yuan_low-rank_2012}\\
\dsat, \dsft& -&-\\
\parbox{9em}{\thetaomegahistogram,\\ \thetacumhisto} & -& \parbox{5em}{EdgeRemoval~\cite{blocki_differentially_2013},\\
Truncation~\cite{kasiviswanathan_analyzing_2013},\\
FlowGraph~\cite{raskhodnikova_efficient_2015}}\\
%%%%%%%%%Outlier-HistoPub & P-HP & GS\\
\bpm& - & EM~\cite{mcsherry_mechanism_2007}, Binary RR~\cite{duchi_local_2013, erlingsson_rappor_2014}\\
\privtree& \priveletstar& \parbox{14em}{UG~\cite{qardaji_differentially_2013,qardaji_understanding_2013,su_differentially_2016}, AG~\cite{qardaji_differentially_2013}, \\Hierarchy~\cite{qardaji_understanding_2013}, DAWA~\cite{li_data-_2014}}\\
\dpcocgen& \privbayes & -\\
\sortaki&- &-\\
\pythia, \delphi& -&-\\
\tru, \minalg, \opt & \boost & \parbox{14em}{n-grams~\cite{chen_differentially_2012}, FreqItem~\cite{zeng_differentially_2012},\\GS, DAWA, DPT~\cite{he_dpt_2015}}\\
\dppro& - & \parbox{9em}{Private SVM~\cite{rubinstein_learning_2012}, \\
PriView~\cite{qardaji_priview_2014},\\ JTree~\cite{chen_differentially_2015}} \\
\tlambda & - &-\\
\gga& \dsat&-\\
\prish & - & MWEM, DAWA \\
\ihp, \mihp & \boost,\efpa, \php, \structurefirst, \ahp & PSD, GS\\
\rcf & \boost, \noisefirst & SHP~\cite{bassily_local_2015}\\
\end{tabularx}
    \caption{Algorithms used in empirical comparisons, divided by internal (included in the SLR) and external (excluded from the SLR) algorithms, sorted by year of publication. Comparisons with the Laplace mechanism and the author's own defined baselines (such as optimal) have been excluded from the table.}
    \label{tab:comparison-matrix}
\end{table}

\FloatBarrier

\subsection{Categorization of Differentially Private Accuracy Improving Techniques}\label{sec:dimensions}
We observe from the algorithms in the 27 papers, there are three different dimensions to accuracy improvement in the context of differential privacy: \textbf{i)} \textit{total noise reduction}, \textbf{ii)} \textit{sensitivity reduction} and \textbf{iii)} \textit{dimensionality reduction}.

\begin{enumerate}
    \item[i)]{\textbf{ Total Noise Reduction}} On the one hand, a histogram is published as statistical representation of a given \dataset\ (Goal I). On the other hand, histograms are published as a way to approximate the underlying distribution, which is then used to answer queries on the \dataset\ (Goal II). We refer to the latter as universal histograms: terminology adapted from ~\cite{hay_boosting_2010}. In this dimension, optimizing the noisy end result (i.e differentially private histogram) provides opportunities for accuracy improvement. 
\item[ii)]{\textbf{Sensitivity Reduction}} The global sensitivity of histogram queries is not small for graph \dataset s. Because, even a \textit{relatively} small change in the network structure results in big change in the query answer. The accuracy improvement in this dimension follow from global sensitivity optimization. 
\item[iii)]{\textbf{Dimensionality Reduction}} Publishing synthetic version of an entire \dataset\ consists of building a private statistical model from the original \dataset\ and then sampling data points from the model. In this dimension, inferring the underlying data distribution from a smaller set of attributes provides opportunities for accuracy improvement.
\end{enumerate}

\subsubsection{Dimension: Total Noise Reduction}\label{dim:noisereduce}

In \Cref{tab:goalnoise}, we summarize the distinct techniques/approaches of the state-of-the-art from the point of view of reducing the total noise.

\begin{table}[htb]\footnotesize
    \centering %\rowcolors{1}{}{gray!10}
    \begin{tabularx}{\linewidth}{>{\bfseries}KlrZ}
        \toprule
        Category & \textbf{Technique/Approach} & \textbf{Algorithms} & \textbf{Notes} \\
        \midrule
        {\multirow{16}{*}{Clustering}} 
         & Bi-partite & \bpm &\\
         & Bisection & \php &\\
         & Bisection & \ihp, \mihp &\\
         & MODL co-clustering~\cite{boulle_data_2011}& \dpcocgen &\\
         & Matrix decomposition & \priveletplus &\\
         & Weighted combination & \ac & Least Square Method\\
         & Retroactive Grouping & \rg & Thresholded\\
         & Selecting Top \textit{k} & \efpa &  \\
         & & \citm & Key/foreign-key Relationships\\
         &  & \minalg & Query Overlap\\
         &  & \structurefirst& V-optimality\\
         &  & \noisefirst & \\
         &  & \ahp& V-optimality\\
         &  & \thetaomegahistogram & V-optimality\\
         &  & \tlambda & Equi-width\\
        \midrule
        {\multirow{8}{*}{\parbox{6em}{\begin{flushright}Consistency\\ Check\end{flushright}}}} & Frequency Calibration &\thetacumhisto & Monotonicity Property\\
            & Hierarchical Consistency & \opt & \\
            &Least Square Minimization & \boost &\\
            &Least Square Minimization & \dpcube & \\
            &Realizable model & \citm & Linear-time Approximation \\
            & & \pmost & Least Norm Problem\\
        \midrule
        {\multirow{6}{*}{\parbox{7em}{\begin{flushright}Hierarchical\\ Decomposition\end{flushright}}}} & Binary Tree & \boost\\
        & kd-tree & \dpcube & V-optimality\\
        & Quadtree & \privtree &\\
        & Query Tree & \citm & Correlation of $i$-Table Model\\
        & Sequential Partitions& \mihp & t-value\\
        \midrule
        {\multirow{6}{*}{\parbox{6em}{\begin{flushright}Learning\\ True \\Distribution\end{flushright}}}} & Reallocate Values & \thetacumhisto & Linear Regression, Powerlaw \& Uniform distributions\\
        & Rescaling Weights & \prish & Query Absolute Error, Dependency Constraints\\
        \midrule
        {\multirow{8}{*}{\parbox{6em}{\begin{flushright}Privacy\\Budget\\Optimization\end{flushright}}}} & Composition rule-based & \citm &  \\
        &  Threshold-driven Release& \dsat, \dsft&  Adaptive-distance Qualifier, Fixed-distance Qualifier\\
        &  Threshold-driven Release & \gga & Fixed-distance Qualifier\\
        &   Weighted & \bpm & \\
        \midrule
        {\multirow{2}{*}{Sampling}} &  Bernoulli Sampling & \rg &\\
        &  Data Recycling & \drpp &\\
        \midrule
        {\multirow{1}{*}{Sorting}} &   & \ahp & \\
        \midrule
        {\multirow{2}{*}{Transformation}} & Wavelet Transform & \privelet &\\
        & Fourier Transformation& \efpa&\\
        \midrule
        {\multirow{5}{*}{Threshold}} & Qualifying Weight & \pmost &\\
        & Qualifying Source-of-noise & \bmax &\\
        & Qualifying Source-of-noise & \tru &\\
        & Sanitization & \ahp &\\
        &  Wavelet Thresholding & \priveletstar &\\
        \bottomrule
    \end{tabularx}
    \caption{Categorization of techniques/approaches used by each algorithm for total noise reduction. Additional qualifiers of each techniques are captured as notes.}\label{tab:goalnoise}
\end{table}

\begin{enumerate}
    \item[$\triangleright$]\textbf{Goal I:} When the goal is to publish some statistical summary of a given \dataset\ as a differentially private histogram, histogram partitions play an essential role in improving the accuracy of the end result. A histogram partitioned into finer bins reduces approximation error\footnote{Error caused by approximating the underlying distribution of data into histogram bins: intervals covering the range of domain values.} of the result, because each data point is correctly represented by a bin. However, the Laplace mechanism for histograms adds noise of scale $\Delta f/\mathepsilon$ to each histogram bin. In other words, a histogram that is structured to minimize the approximation error, would suffer more noise in order to satisfy differential privacy.
\end{enumerate}
%since it utilizes the maximum number of bins in order to represent the histogram structure.
The most common approach to enhance the utility for this goal, is to identify optimal histogram partitions for the given data.

Algorithms \php, \structurefirst\ and  \thetaomegahistogram\ use the Exponential mechanism to find V-optimal histogram~\cite{jagadish_optimal_1998} partitions. However, the quality of the partitions drops as the privacy budget available for iterating the Exponential mechanism decreases. Hence, algorithms \noisefirst, \ahp, \dpcocgen\ instead operate on the non-optimized noisy histogram for identifying sub-optimal partitions for the final histogram. To further improve the quality of the partitions that are based on the non-optimized noisy histogram, in \ahp sorting technique is used.

For the same goal described above, if the given data are bitmap strings then one opportunity for accuracy improvement is to vary the amount of noise for various histogram bins. Algorithm \bpm\ uses a bi-partite cut approach to partition a weighted histogram into bins with high average weight and bins with low relative weight. Further, in \bpm\ the privacy budget \ep\ is carefully split between the bins such that the heavy hitters enjoy less noise. Algorithm \ac\ uses weighted combination approach in terms of least square method in order to find optimal histogram partitions. Sample expansion through recycling the data points is another interesting approach for enhancing the accuracy of histogram over bitmap strings.  

In the case of dynamic \dataset s, it is desirable to sequentially release the statistical summary of evolving \dataset\ at a given point in time. The most common approach is to limit the release of histograms, when there is a change in the \dataset\ for avoiding early depletion of privacy budget. Algorithms \dsft, \dsat\ and \gga\ uses distance based sampling to monitor significant updates to the input \dataset. In algorithm \rg\ an adaptive sampling process uses Bernoulli sampling for change detection in the \dataset. Further, in \rg\ a novel histogram partitioning approach called retroactive grouping is introduced to enhance the accuracy of the end result.

\begin{enumerate}
    \item[$\triangleright$]\textbf{Goal II:} When the histograms are used to answer workload of allowable queries. Laplace noise accumulates (sequential composition) as the number of queried histogram bins increases in order to answer the workload (covering large ranges of domain values). However, if the answer to the workload can be constructed by finding a linear combination of fewer bins, then the accuracy of the final answer will be significantly improved.
\end{enumerate}
Algorithms \boost, \dpcube, \privtree, \citm\ and \mihp\ employ an approach, where the domain ranges are hierarchically structured, typically in a tree structure. The intuition is, to find the fewest number of internal nodes such that the union of these ranges equals the desired range in the workload. To further improve the accuracy in the context of sequential composition, algorithm \citm\ uses composition rule-based privacy budget optimization. Transformation techniques such as wavelet transform (\privelet) and Fourier transform (\efpa) are also used to model linear combination of domain ranges.

Another approach to reduce the accumulate noise in the context of universal histogram is to contain the total noise below a threshold. In \bmax\ the maximum noise variance of the end result is contained within a threshold. 

Furthermore, constraints are imposed in the output space of possible answers, which are then verified in the \postprocessing\ step to identify more accurate answers in the output space.

Preserving the dependency constraint is important for answering range queries over spatial histogram. To this end, in algorithm \prish, true distribution of the underlying \dataset\ is learned from private answers to carefully chosen informative queries. Separately, to estimate the tail distribution of the final noisy histogram, algorithm \thetacumhisto\ uses some prior distribution to reallocate count values.

\FloatBarrier

\subsubsection{Dimension: Sensitivity Reduction}\label{dim:sensreduce}

In \Cref{tab:goalsensitivity}, we summarize the distinct techniques/approaches of the state-of-the-art from the point of view of reducing the global sensitivity.

\begin{table}[htb]
    \centering
    \begin{tabularx}{\linewidth}{>{\bfseries}KlrZ}
        \toprule
        Category & \textbf{Technique/Approach} & \textbf{Algorithms} & \textbf{Notes} \\
        \midrule
        Neighbor Relation & Redefine & \citm & Propagation Constraints\\
        \midrule
        {\multirow{3}{*}{Projection}} & Edge Addition & \parbox{8em}{{\begin{flushright} \thetaomegahistogram\\ \thetacumhisto\end{flushright}}} & Network Degree Bounded\\
        & Edge Deletion & \tlambda& Mutual Connections Bounded\\
        \bottomrule
    \end{tabularx}
    \caption{Categorization of techniques/approaches used by each algorithms for sensitivity reduction. Additional qualifiers of each techniques are captured as notes.}
    \label{tab:goalsensitivity}
\end{table}

In graph \dataset s, global sensitivity becomes unbounded, for example, change in a node and its edges, in the worst case affects the whole structure (i.e involving all the nodes) of the network under \textit{node differential privacy}. Bounding the network degree is one of the common approaches for containing the global sensitivity for analysis under \textit{node differential privacy}. Techniques, edge addition (\thetaomegahistogram, \thetacumhisto) and edge deletion (\tlambda) are used to bound the size of the graph. Consequently, the noise required to satisfy \textit{node differential privacy} will be reduced.

When there exists no \textit{standard} neighborhood definition for the differential privacy guarantee in the light of correlated data structures. In the \citm\ algorithm that operates on relational databases with multiple relation correlations, the neighbor relation is redefined.

\subsubsection{Dimension: Dimensionality Reduction}\label{dim:dimreduce}

In \Cref{tab:goaldim}, we summarize the distinct techniques/approaches of the state-of-the-art from the point of view of reducing the data dimensions.

\begin{table}[htb]
    \centering %\rowcolors{1}{}{gray!10}
    \begin{tabularx}{\linewidth}{>{\bfseries}KlrZ}
        \toprule
        Category & \textbf{Technique/Approach} & \textbf{Algorithms} & \textbf{Notes} \\
        \midrule
        Consistency check &  Eigenvalue Procedure~\cite{rousseeuw_transformation_1993} & \dpcopula &\\
        \midrule
        Projection &  Hashing Trick~\cite{weinberger_feature_2009} & \pegs &\\
        \midrule
        Privacy Budget Optimization & Reset-then-sample  & \pegsrs &\\
        \midrule
        {\multirow{3}{*}{Transformation}} & Bayesian Network & \privbayes &\\
        & Copula Functions & \dpcopula &\\
        & Random Projection & \dppro & Johnson-Lindenstrauss Lemma\\
        \bottomrule
    \end{tabularx}
    \caption{Categorization of techniques/approaches used by each algorithm for data dimensionality reduction. Additional qualifiers of each techniques are captured as notes.}
    \label{tab:goaldim}
\end{table}

The most common approach to accuracy improvement in this dimension is to build statistical models that approximate the full dimensional distribution of the \dataset\ from multiple set marginal distributions. Some of techniques to approximate joint distribution of a \dataset\ are Bayesian Network (\privbayes) and Copula functions (\dpcopula). Furthermore, projection techniques from high-dimensional space to low-dimensional sub-spaces are shown to improve accuracy as less noise is required to make the smaller set of low-dimensional sub-spaces differentially private. Projection techniques found in the literature are, feature hashing using the hashing trick (\pegs) and random projection based on the Johnson-Lindenstrauss Lemma (\dppro).

In \dpcopula, eigenvalue procedure is used in the \postprocessing\ stage to achieve additional gain in accuracy. Unexpectedly, reset-then-sample approach grouped under privacy budget optimization algorithmic category appear in this dimension, because the \pegsrs\ algorithm supports multiple synthetic \dataset\ instances.

\subsubsection{Summary}
\begin{figure}[htb]
    \centering
    \includegraphics[width=\linewidth]{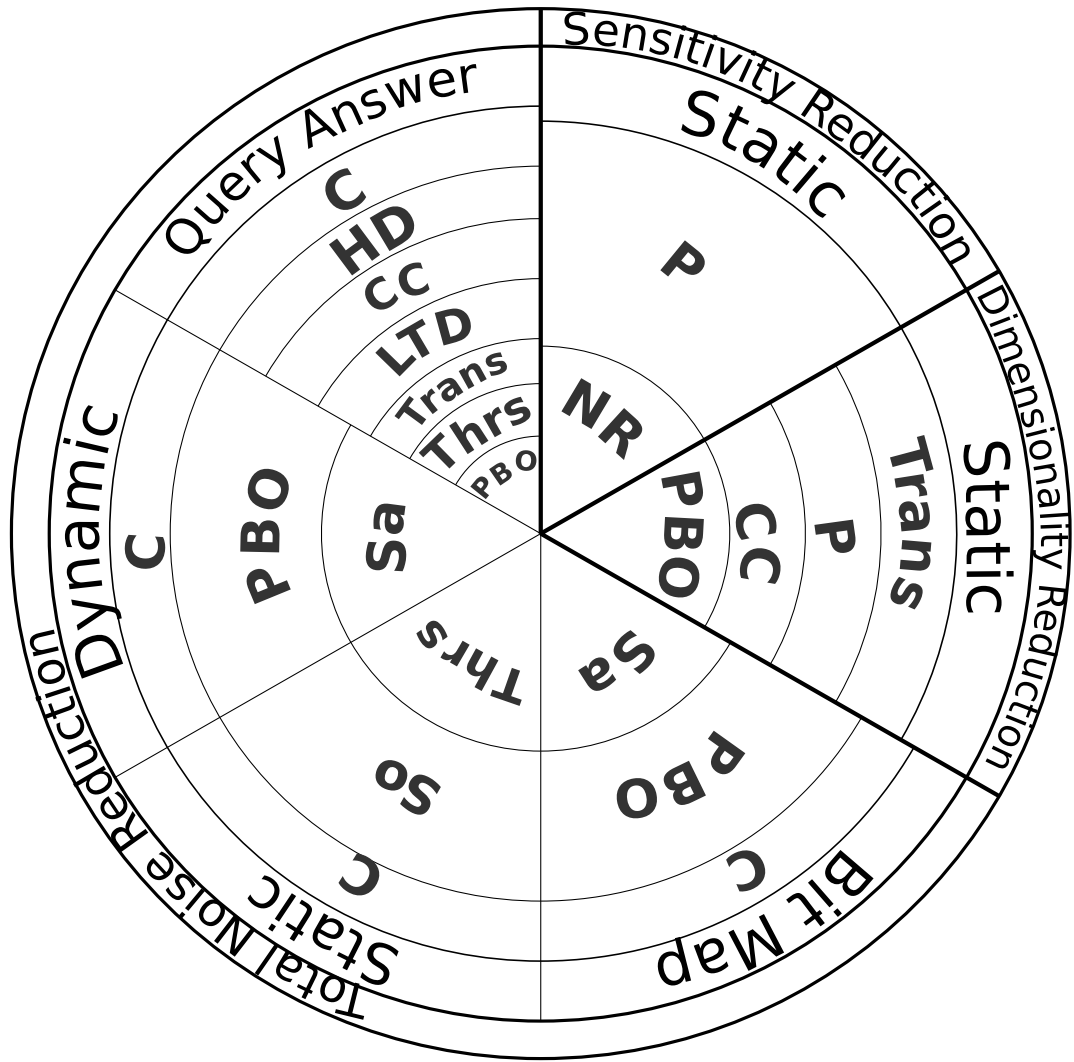}
    \caption{Conceptualization of accuracy improving techniques in the context of differential privacy: Abbreviations: {\textbf{C}}: Clustering, {\textbf{CC}}: Consistency Check, {\textbf{HD}}: Hierarchical Decomposition, {\textbf{LTD}}: Learning True Distribution, {\textbf{NR}}: Neighborhood Redefine, {\textbf{P}}: Projection, {\textbf{PBO}}: Privacy Budget Optimization, {\textbf{Thrs}}: Threshold, {\textbf{Trans}}: Transformation, {\textbf{Sa}}: Sampling, {\textbf{So}}: Sorting}
    \label{fig:goals-view}
\end{figure}

\Cref{fig:goals-view} summarizes the categorization of differentially private accuracy improving techniques. Techniques identified in each accuracy improving dimensions are grouped into specific categories. The algorithmic categories are further partially sub-divided by the input data they support. Query answer relates to the type of release rather than to the input data, but the assumption is that the other mentioned data types, they implicitly specify the type of release.

A further located the algorithmic category from the center of the circle, the more common is that category in that particular accuracy improvement dimension. Subsequently, clustering is the most commonly employed category for the total noise reduction dimension. Interestingly, same set of categories of accuracy improving algorithms are employed for dynamic data and bitmap strings, in the context of total noise reduction dimension. Hierarchical decomposition, consistency check and learning true distribution are primarily used in the context of releasing histogram for answering workload of queries. It should be noted that the consistency check technique is used in the dimensionality reduction dimension as well but the usage of the technique is conditional. 

% consistency check and trans and bpo change

\FloatBarrier
\section{Discussion}\label{sec:discussion}

One limitation of this paper is that the scope of our SLR is limited to papers with empirical results. We have chosen empirical measurement of accuracy, since it can provide a less pessimistic understanding of error bounds, as opposed to analytical bounds. However, in our analysis (\Cref{sec:analysis}) of the papers, we studied related theoretical aspects of accuracy improvements and put the surveyed papers into context by tracing their origin, illustrated in \Cref{fig:family-tree}. As such, we can guide the interested reader in the right direction, but we do not provide an analysis of theoretical results.

%i) some algorithms work best for some type of data or if the \dataset\ follows a particular data distribution ii) tuning different algorithmic parameters has implications in terms of privacy guarantee or accuracy, iii) computational complexity  for different workloads. 

\subsection{Composability of Categories}\label{sec:composability}
%Most common group of techniques are trunction, hierarchical decomposition and post processing. Finish the dimensionality of papers diagram and complete this section. How truction can be integrated in ---- techniques but will cost in terms of accuracy and/or privacy or something else?

%{\footnotesize{SORTaki}}

From the dimensions identified in our analysis, we continue by investigating how techniques from different categories \textit{may} be composed. We also connect the papers with the \textit{place}~\footnote{Places refers to different points in the workflow of a typical differentially private analysis, see \Cref{fig:improvement-places}} their algorithm operates on in \Cref{tab:composition}. We believe a technique from one place is possible to compose with techniques from another place, since the places are designed to be a sequential representation of the data analysis. We will illustrate this composability by giving a few examples of how techniques are composed in the included papers.

\begin{table}[H]
\centering\rowcolors{2}{}{gray!10}\resizebox{0.3\paperheight}{!}{
\begin{tabularx}{0.7\linewidth}{rp{2em}p{2em}p{2em}p{2em}}
 \cellcolor{white}& \textbf{A} & \textbf{B} & \textbf{C} & \textbf{D} \\ 
\citet{hay_boosting_2010} & & \x & & \\
\citet{ding_differentially_2011} & & \x & \x & \\
\citet{xiao_differential_2011} & & & & \x \\
\citet{acs_differentially_2012} & & & \x & \x \\ 
\citet{xu_differentially_2013} & & & \x & \\
\citet{li_differentially_2014} & & & \x & \\ 
\citet{lu_generating_2014} & & \x & & \x \\ 
\citet{park_pegs_2014} & & \x & & \x \\  
\citet{xiao_dpcube_2014} & & & \x & \\  
\citet{zhang_privbayes_2014} & & &\x & \\  
\citet{zhang_towards_2014} & & &\x & \\ 
\citet{chen_private_2015} & & \x & & \\  
\citet{lee_maximum_2015} & & \x &  & \\
\citet{li_differentially_2015} & \x & & \x & \\
\citet{day_publishing_2016} &\x &\x &\x & \\ 
\citet{wang_private_2016} & & &\x &\\
\citet{zhang_privtree_2016} & & &\x & \\ 
\citet{benkhelif_co-clustering_2017} & & \x & \x & \\ 
\citet{doudalis_sortaki_2017} & & &\x & \\ 
\citet{kotsogiannis_pythia_2017} &\x &\x & &\x \\
\citet{wang_differentially_2017} & & & \x & \\ 
\citet{xu_dppro_2017} & & & \x & \\
\citet{ding_privacy-preserving_2018} &\x & & & \x \\ 
\citet{gao_dynamic_2018} & & & \x & \\  
\citet{ghane_publishing_2018} & &\x &\x & \\
\citet{li_ihp_2019} & & & \x & \\ 
\citet{nie_utility-optimized_2019} & & \x & & \x \\  
\end{tabularx}}
\caption{Mapping the papers to each place where: A) Altering the query, B) Post-processing, C) Change in mechanism, D) Pre-processing}
\label{tab:composition}
\end{table}

\subsubsection*{Place A: Altering the Query}
Altering the query targets \textit{sensitivity reduction}, as sensitivity is a property of the query. Our take away from the SLR is that there are mainly two tricks to altering the query:
\begin{enumerate}
    \item\label{item:break-down-query} When an analysis requires a high sensitivity query, replace the query with an approximate query, or break the query down into two or more sub-queries.
    \item\label{item:sampling} Use sampling to avoid prematurely exhausting the privacy budget.
\end{enumerate}

\textbf{\Cref{item:break-down-query}:} For example, a histogram query is broken down into two separate queries: a clustering technique based on the exponential mechanism and usually a Laplace counting query, as in the case with \citet{xu_differentially_2013} and consecutive work.

By breaking down the query, the sensitivity reduction can increase accuracy, but it needs to be balanced against the source of accumulated noise that is introduced by multiple queries. In particular, when breaking down a query, the privacy budget needs to be appropriately distributed between the sub-queries. For example, when breaking a histogram into a clustering query and then a count query, one could choose to give more budget to the clustering step to find a tighter histogram structure, but that would come at the cost of less accuracy for the count query.

\textbf{\Cref{item:sampling}:} When an analysis in done on dynamic data, it is possible to unintentionally include the same data points in multiple queries, and ending up 'paying' for them multiple times. \citet{li_differentially_2015} mitigates this source of accumulated noise by deploying sampling. It is also possible to use sampling for static data, for example, \delphi\ by \citet{kotsogiannis_pythia_2017} could be trained on a sample of the full \dataset, if no public training data is available.

\subsubsection*{Place B: \Postprocessing}
\Postprocessing\ targets \textit{total noise reduction}, usually by exploiting consistency checks or other known constraints. Since \postprocessing\ is done on data that has been released by a differentially private algorithm, \postprocessing\ can always be done without increasing the privacy loss. However, \postprocessing\ can still decrease accuracy if used carelessly. In our SLR, the main \postprocessing\ idea is:
\begin{enumerate}
    \item\label{item:boost} Finding approximate solutions to get rid of inconsistencies through \textit{constrained inference}~\cite{hay_boosting_2010}.
    \item\label{item:maximumlikelihood} Applying consistency checks that would hold for the raw data.
\end{enumerate}

\textbf{\Cref{item:boost}:} \boost\ is already being combined with several algorithms that release histograms, for example \noisefirst\ and \structurefirst. \admm\ is a similar, but more generic solution that has been applied to more output types than just histograms. In fact, \citet{lee_maximum_2015} claims \admm\ can re-use algorithms use for least square minimization, which means \boost\ should be possible to incorporate in \admm. Consequently, we believe \admm\ would compose with most algorithms due to its generic nature.

\subsubsection*{Place C: Change in the Release Mechanism}
Changing the release mechanism mainly targets \textit{total noise reduction}. In the SLR, we found the following approaches being used:
\begin{enumerate}
    \item\label{item:test-and-release} Test-and-release.
    \item\label{item:sorting} Sorting as an intermediary step.
\end{enumerate}

\textbf{\Cref{item:test-and-release}:} \dsat\ and \dsft\ uses thresholding to determine when to release data, as a way to save the privacy budget. Thresholding is particularly useful for dynamic data, as it often requires multiple releases over time. For example, adaptive or fixed thresholding can be used for sensor data and trajectory data, effectively providing a way of sampling the data. 

\structurefirst\ also uses a type of test-and-release when creating the histogram structure using the exponential mechanism. The test-and-release approach means EM can be combined with basically any other release mechanism, which is also what we found in the literature. We believe the main challenge with EM is finding an adequate scoring/utility function, and this is where we believe a lot of accuracy improvement will come from.

\Cref{item:sorting} \sortaki\ is designed to be composable with two-step algorithms that release histograms, for example \noisefirst. The idea is that by sorting noisy values, they can group together similar values that would otherwise not be grouped due to the bins not being adjacent.

\subsubsection*{Place D: \Preprocessing}
\Preprocessing\ generally targets \textit{dimensionality reduction} or \textit{total noise reduction}. In our SLR, we encountered the following types of \preprocessing:
\begin{enumerate}
    \item\label{item:projection} Encoding through projection/transformation.
    \item\label{item:nonsensitive} Learning on non-sensitive data.
\end{enumerate}

\textbf{\Cref{item:projection}:} Several algorithms project or transform their data, for example \privelet\ and \efpa. Encoding can reduce both sensitivity and dimensionality by decreasing redundancy, and is therefore especially interesting for multi-dimensional as well as high-dimensional, sparse, \dataset s. However, lossy compression techniques can potentially introduce new sources of noise, and therefore adds another \tradeoff\ that needs to be taken into account. Intuitively, lossy compression is beneficial when the noise lost in the compression step is greater than the proportion of useful data points lost. For example, sparse data may benefit more from lossy compression than data that is not sparse.

\textbf{\Cref{item:nonsensitive}:} \delphi\ is a \preprocessing\ step which uses a non-sensitive, public \dataset\ to build a decision tree. In cases where public \dataset s are available, it could be possible to adopt the same idea; for example learning a histogram structure on public data as opposed to spending budget on it. The caveat here is of course that the public data needs to be similar enough to the data used in the differentially private analysis, because otherwise this becomes an added source of noise. Thus, learning from non-sensitive data introduces another \tradeoff\ that is still largely unexplored.

\subsection{Incomparable papers}\label{sec:incomparable}
We present a list of papers that were excluded during our qualitative analysis, and the reason for why we decided to exclude them in \Cref{sec:analysis}. The reason for excluding papers in the analysis step is that certain properties of their algorithms make them incomparable with other algorithms.

\begin{enumerate}
\item[\cite{yan_differentially_2018}:] The DP-FC algorithm does not consider the structure of the histogram a sensitive attribute, and thus achieves a trivial accuracy improvement over other algorithms.
\item[\cite{liu_histogram_2018}:] The APG algorithm does not perform differentially private clustering, and therefore achieves better accuracy by relaxing the privacy guarantees compared to \ahp, \ihp\ and GS.
\item[\cite{qian_publishing_2018}:] The SC algorithm uses the ordering of the bins in order to calculate the cluster centers, but does not perturb the values before doing so, and thus the order is not protected, making their guarantees incomparable.
\item[\cite{han_publishing_2016}:] The Outlier-Histopub algorithm, similarly sorts the bin counts according to size, without using the privacy budget accordingly to learn this information. The authors claim that this type of sorting does not violate differential privacy, but due to the fact that the output is determined based on the private data, the approach cannot be 0-differentially private.
\item[\cite{li_research_2018}:] The ASDP-HPA algorithm does not describe the details of how their use of Autoregressive Integrated Moving Average Model (ARIMA) is made private, and thus we cannot determine whether the entire algorithm is differentially private. Furthermore, the details of how they pre-process their \dataset\ is not divulged, and it can thus not be determined if the \preprocessing\ violates differential privacy or not by changing the query sensitivity.
\item[\cite{hadian_privacy-preserving_2016}:] The algorithm is incomplete, since it only covers the histogram partitioning, and does not involve the addition of noise to bins. Furthermore, it is not clear whether they draw noise twice using the same budget, or if they reuse the same noise for their thresholds. As the privacy guarantee \ep\ cannot be unambiguously deduced, we do not include their paper in our comparison.
\item[\cite{chen_iterative_2013}:] The GBLUE algorithm generates a k-range tree based on the private data, where k is the fanout of the tree. Since private data is used to decide on whether a node is further spilt or not, it does not provide the same privacy guarantees as the other studied algorithms. 
\item[\cite{li_differential_2017}:] The algorithm creates groups based on the condition that the merged bins guarantee $k$-indistinguishability. Since this merge condition is based on the property of the data it does not guarantee differential privacy on the same level as the other papers, so we deem it incomparable.
\end{enumerate}

Further, in the analysis regarding dimensions of accuracy improvement techniques presented in \Cref{sec:analysis}, some algorithms such as \admm, \sortaki\ and \pythia\ are excluded. The rationale behind the exclusion is, these algorithms are not self contained, but nevertheless improves accuracy of the differentially private answers when combined with other analyzed algorithms.

Efforts such as \pythia\ and {\footnotesize{DPBench}}~\cite{hay_principled_2016}, that provide practitioners a way to empirically assess the privacy-accuracy \tradeoff\ related to their \dataset s are commendable. However, to effectively use the tool one needs to have some background knowledge of the right combination of parameters to tune. In our analysis of the algorithms, we mapped out the accuracy improvement techniques grouped by optimization goals and corresponding query size. This knowledge will allow practitioners and researchers alike to think about other places to explore for accuracy improvement, rather than finding the algorithms that are based only on their data. Essentially, we provide an understanding to enable algorithm design, as opposed to algorithm selection.

%Limitations in the method? Limitations in the analysis?

%Some papers have no connections, which no connections internally can be due to the fact that the SLR only includes papers with empirical results, and papers' connections to more theoretical papers will thus be shown in the external positioning.
\section{Conclusions}\label{sec:conclusion}
Motivated by scarcity of works that structure knowledge concerning accuracy improvement in differentially private computations, we conducted a systematic literature review (SLR) on accuracy improvement techniques for histogram and synthetic data publication under differential privacy.

We present two results from our analysis that addresses our research objective, namely to synthesize the understanding of the underlying foundations of the privacy-accuracy \tradeoff\ in differentially private computations. This systematization of knowledge (SoK) includes:
\begin{enumerate}
    \item Internal/external positioning of the studied algorithms (\Cref{fig:family-tree} and \Cref{tab:comparison-matrix}).
    \item A taxonomy of different \textit{categories} (\Cref{fig:goals-view}) and their corresponding \textit{optimization goals} to achieve accuracy improvement: \textit{total noise reduction} (\Cref{tab:goalnoise}), \textit{sensitivity reduction} (\Cref{tab:goalsensitivity}) and \textit{data dimensionality reduction} (\Cref{tab:goaldim}).
\end{enumerate}
What's more, we also discuss and present an overview of composable algorithms according to their optimization goals and category, sort-out by the \textit{places}, in which they operate (\Cref{sec:composability}). Our intent is that these findings will pave the way for future research by allowing others to integrate new solutions according to the categories. For example, our places can be used to reason about where to plug in new or existing techniques targeting a desired \textit{optimization goal} during algorithm design.

%Furthermore, we elaborate on research problems that arise from moving the original challenge, which also opens up for future research.
% We believe from our comprehensible categorization of the different approaches, that, it is possible to take ideas from other disciplines and plug them in as new techniques, which opens the door for many possible research directions. 

From our overview of composability, we see that most efforts are focused on making \textit{changes in the mechanism}, and on \textit{\postprocessing}. We observe that, \textit{altering the query} in one way or another, is not popular, and we believe further investigation is required to understand which techniques can be adopted in this place.

Finally, although all algorithms focus on accuracy improvement, it is impossible to select the 'best' algorithm without context. Intuitively, newer algorithms will have improved some property of an older algorithm, meaning that newer algorithms \textit{may} provide higher accuracy. Still, the algorithms are used for different analyses, which means not all algorithms will be interchangeable. Secondly, many algorithms are data dependent, which means that the selection of the 'best' algorithm may change depending on the input data used, even when the analysis is fixed. Consequently, the 'best' algorithm needs to be chosen with a given \dataset\ and a given analysis in mind. The problem of choosing the 'best' algorithm when the setting is known is in fact addressed by \pythia.

% \section*{Acknowledgements}
\mdseries

\printbibliography

\begin{appendices}
\newpage
\begin{appendix}\appendix
\crefalias{section}{appsec}
\section{Excluded Papers}\label{app:appendix}
\subsection*{Query 1}

\begin{longtable}{|l|c|}
\caption{Excluded papers from query 1 (focusing on histograms), and the corresponding exclusion criteria}\\
\hline
\textbf{Citation} & \textbf{Exclusion Criteria}\\ \hline
\endfirsthead
\multicolumn{2}{c}%
{\tablename\ \thetable\ -- \textit{Continued from previous page}} \\
\hline
\textbf{Citation} & \textbf{Exclusion Criteria}\\
\hline
\endhead
\hline \multicolumn{2}{r}{\textit{Continued on next page}} \\
\endfoot
\hline
\endlastfoot
\citet{balcer_differential_2018} & 2\\ \hline
\citet{bassily_local_2015} & 6\\ \hline
\citet{benkhelif_publication_2018} & 9, 10\\ \hline
\citet{bhowmick_differential_2018} & 7\\ \hline
\citet{bowen_differentially_2018} & 8 \\ \hline
\citet{bowen_statistical_2018} & 8\\ \hline
\citet{chaudhuri_stability-based_2013} & 5 \\ \hline
\citet{cyphers_anonml_2017} & 5 \\ \hline
\citet{eugenio_cipher_2018} & 5\\ \hline
\citet{fanaeepour_case_2015} & 1\\ \hline
\citet{fanaeepour_end--end_2017} & 4 \\ \hline
\citet{fanaeepour_histogramming_2018} & 2 \\ \hline
\citet{fei_differential_2018} & 7 \\ \hline
\citet{foote_releasing_2019} & 5\\ \hline
\citet{gardner_share_2013} &2 \\ \hline
\citet{gehrke_crowd-blending_2012} &3 \\ \hline
\citet{hall_random_2013} &3 \\ \hline
\citet{hardt_multiplicative_2010} & 1, 2, 6 \\ \hline
\citet{hardt_geometry_2010} & 6 \\ \hline
\citet{kellaris_engineering_2019} & 2 \\ \hline
\citet{kobliner_locally_2018} & 7\\ \hline
\citet{kulkarni_answering_2018} & 9 \\ \hline
\citet{lan_greedy_2013} & 10 \\ \hline
\citet{lei_differentially_2011} & 5\\ \hline
\citet{li_optimizing_2010} & 6\\ \hline
\citet{li_lit_survey_2015} & 2 \\ \hline
\citet{li_efficient_2015} & 2\\ \hline
\citet{li_privacy_2018} & 2\\ \hline
\citet{li_differentially_2019} & 5\\ \hline
\citet{li_impact_2019} & 1, 2, 5\\ \hline
\citet{lin_information_2013} & 1, 2, 5 \\ \hline
\citet{ling_detrended_2018} & 7 \\ \hline
\citet{luo_predictable_2019} & 1, 2, 3 \\ \hline
\citet{meng_different_2017} & 2 \\ \hline
\citet{naghizade_challenges_2017} & 1 \\ \hline
\citet{nikolov_geometry_2016} & 6 \\ \hline
\citet{raigoza_differential_2017} & 2, 6, 9 \\ \hline
\citet{roth_differential_2010} & 6\\ \hline
\citet{shang_application_2014} & 2\\ \hline
\citet{smith_more_2017} & 1\\ \hline
\citet{su_privpfc_2018} & 5\\ \hline
\citet{xiao_ireduct_2011} & 3\\ \hline
\citet{xiaoling_histogram-based_2015} & 7\\ \hline
\citet{ying_linear_2013} & 2, 6\\ \hline
\citet{zhang_differentially_2017-1} & 7\\ \hline
\citet{zhu_correlated_2015} & 2\\ \hline
\end{longtable}
    \label{tab:query1-full}

\newpage

\subsection*{Query 2}

\begin{longtable}{|l|c|}
\caption{Excluded papers from query 2 (focusing on synthetic data), and the corresponding exclusion criteria} \\
\hline
\textbf{Citation} & \textbf{Exclusion Criteria}\\
\hline
\endfirsthead
\multicolumn{2}{c}%
{\tablename\ \thetable\ -- \textit{Continued from previous page}} \\
\hline
\textbf{Citation} & \textbf{Exclusion Criteria}\\
\hline
\endhead
\hline \multicolumn{2}{r}{\textit{Continued on next page}} \\
\endfoot
\hline
\endlastfoot
\citet{abay_privacy_2018} & 2 \\ \hline
\citet{abowd_how_2008} & 1 \\ \hline
\citet{aliakbarpour_differentially_2017} & 1 \\ \hline
\citet{balog_differentially_2018} & 2, 6\\ \hline
\citet{barak_privacy_2007} & 6\\ \hline 
\citet{barrientos_providing_2018} & 1\\ \hline 
\citet{barrientos_differentially_2018} & 1\\ \hline
\citet{bindschaedler_plausible_2017} & 3\\ \hline
\citet{blum_learning_2008} & 2\\ \hline
\citet{blum_learning_2013} & 2\\ \hline
\citet{bohler_privacy-preserving_2017} & 4 \\ \hline
\citet{bousquet_passing_2019} & 1\\ \hline
\citet{bowen_comparative_2016} & 2,6 \\ \hline 
\citet{bowen_differentially_2016} & 8\\ \hline
\citet{bowen_statistical_2018} & 8\\ \hline
\citet{cao_quantifying_2017} &1, 2 \\ \hline
\citet{cao_priste_2018} & 1\\ \hline
\citet{charest_how_2011} & 1 \\ \hline
\citet{chen_wavecluster_2015} & 1\\ \hline
\citet{cormode_constrained_2018} & 1, 2 \\ \hline
\citet{dwork_complexity_2009} & 2, 6 \\ \hline
\citet{elliot_empirical_2014} & 8 \\ \hline
\citet{fan_adaptive_2014} & 1\\ \hline
\citet{fan_adaptively_2012} & 8 \\ \hline
\citet{garfinkel_-identifying_2016-2} & 1\\ \hline
\citet{garfinkel_-identifying_2016} & 1\\ \hline
\citet{gehrke_crowd-blending_2012} & 2\\ \hline
\citet{gupta_iterative_2012} & 2\\ \hline
\citet{hardt_study_2011}& 9\\ \hline
\citet{hu_privacy-preserving_2018} & 1\\ \hline
\citet{hu_pscluster_2018} & 1\\ \hline
\citet{jordon_pate-gan_2019} &5 \\ \hline
\citet{jorgensen_conservative_2015-1} &1, 2 \\ \hline
\citet{kifer_towards_2010} & 1\\ \hline
\citet{kulkarni_constrained_2018} & 8\\ \hline
\citet{lee_sketch_2010} & 2\\ \hline
\citet{li_optimal_2013} & 1\\ \hline
\citet{li_dpsynthesizer_2014} & 2\\ \hline
\citet{li_lower_2015} & 1, 2\\ \hline
\citet{li_lit_survey_2015} & 2,6\\ \hline
\citet{li_bayesian_2018} & 4\\ \hline
\citet{li_privacy_2018} & 2\\ \hline
\citet{li_towards_2018} & 4\\ \hline
\citet{liu_model-based_2016} & 1\\ \hline
\citet{liu_optimized_2018} & 2\\ \hline
\citet{lu_poster_2017} & 2\\ \hline
\citet{machanavajjhala_privacy_2008} & 3, 6 \\ \hline
\citet{matthews_data_2011} & 2, 6\\ \hline
\citet{mcclure_differential_2012} & 1\\ \hline
\citet{mcclure_relaxations_2015} & 9\\ \hline
\citet{mulle_privacy-integrated_2015} & 1\\ \hline
\citet{neel_how_2018} & 6\\ \hline
\citet{park_statistical_2016} & 10 \\ \hline
\citet{ping_datasynthesizer_2017} & 2\\ \hline
\citet{rodriguez_privacy-preserving_2018} & 2\\ \hline
\citet{shlomo_statistical_2018} & 1\\ \hline
\citet{snoke_pmse_2018} & 2\\ \hline
\citet{snoke_statistical_2018} & 9\\ \hline
\citet{triastcyn_generating_2018} & 2\\ \hline
\citet{ullman_privacy_2013} & 9\\ \hline
\citet{vilhuber_synthetic_2016} & 1\\ \hline
\citet{wang_protecting_2018} & 1, 3, 5\\ \hline
\citet{wang_data_2019} & 2\\ \hline
\citet{weggenmann_syntf_2018} & 1\\ \hline
\citet{xu_ganobfuscator_2019} & 1\\ \hline
\citet{yu_differentially_2017} & 9\\ \hline
\citet{zhang_algorithms_2016} & 9\\ \hline
\citet{zhang_differentially_2018} & 5\\ \hline
\citet{zhou_differential_2009} & 1\\ \hline
\end{longtable}
\label{tab:query2-full}

\end{appendix}

\end{appendices}

\end{document}